\documentclass[preprint]{aastex}

\newcommand{\kms}{km~s$^{-1}$}

\newcommand{\teff}{$T_{\rm{eff}}$}
\newcommand{\grav}{log($g$)}
\newcommand{\etal}{et al.}
\newcommand{\eqw}{$W_{\lambda}$}

\newcommand{\mystar}{OGLE--2007--BLG--349S}
\newcommand{\moaten}{MOA--2008--BLG--310S}
\newcommand{\moaeleven}{MOA--2008--BLG--311S}
\newcommand{\johnstar}{OGLE--2006--BLG--265S}
\newcommand{\moa}{MOA--2006--BLG--099S}
\newcommand{\bensbystar}{OGLE--2008-BLG--209S}

\begin{document}

\title{Clues to the Metallicity Distribution in the
Galactic Bulge: Abundances in \moaten\ and \moaeleven
\altaffilmark{1}}

\author{Judith G. Cohen\altaffilmark{2}, Ian B. Thompson\altaffilmark{3},
Takahiro Sumi\altaffilmark{4}, Ian Bond\altaffilmark{5},
Andrew Gould\altaffilmark{6}, Jennifer A. Johnson\altaffilmark{7},
Wenjin Huang\altaffilmark{8} \& Greg Burley\altaffilmark{3} }

\altaffiltext{1}{This paper includes data gathered with the 6.5 meter Magellan Telescopes located at
Las Campanas Observatory, Chile.} 

\altaffiltext{2}{Palomar Observatory, Mail Stop 105-24,
California Institute of Technology, Pasadena, Ca., 91125, 
jlc@astro.caltech.edu}

\altaffiltext{3}{Carnegie Observatories of Washington,
813 Santa Barbara Street, Pasadena, Ca. 91101,
ian,burley@ociw.edu}

\altaffiltext{4}{Solar-Terrestrial Environment Laboratory,
Nagoya University, Nagoya, Japan; sumi@stelab.nagoya-u.ac.jp}

\altaffiltext{5}{Institute for Information and Mathematical Sciences,
Massey University, Auckland, New Zealand; I.A.Bond@massey.ac.nz}

\altaffiltext{6}{Department of Astronomy, Ohio State
University, 140 W. 18th Ave., Columbus, OH 43210; 
gould@astronomy.ohio-state.edu and Institute 
d'Astrophysique de Paris, 98bis Blvd Arago, Paris, 75014,
gould@astronomy.ohio-state.edu}

\altaffiltext{7}{Department of Astronomy, Ohio State
University, 140 W. 18th Ave., Columbus, OH 43210; 
jaj@astronomy.ohio-state.edu}

\altaffiltext{8}{Palomar Observatory, Mail Stop 105-24,
California Institute of Technology, Pasadena, Ca., 91125,
current address: University of
Washington, Department of Astronomy, Box 351580, 
Seattle, Washington, 98195-1580, hwenjin@astro.washington.edu}

\begin{abstract}

We present abundance analyses based on high dispersion and high
signal-to-noise ratio Magellan spectra of two highly microlensed Galactic
bulge stars in the region of the main sequence turnoff with \teff $\sim 5650$~K.
We find that \moaten\ has [Fe/H]\footnote{We adopt the usual spectroscopic notations
that [A/B] ~ $\equiv ~ log_{10} (N_A/N_B)_* - log_{10} (N_A/N_B)_{\odot}$, 
and that log$[\epsilon(A)] ~  \equiv ~ log_{10} (N_A/N_H) + 12.00$, for elements
$A$ and $B$.} = +0.41$\pm$0.09~dex and \moaeleven\ has +0.26$\pm$0.09~dex.
The abundance ratios for the $\sim$20 elements for which features could
be detected in the spectra of each of the two stars follow the trends
with [Fe/H] found among samples of bulge giants.
Combining
these two bulge dwarfs with the results from previous abundance analysis of 
four other
Galactic bulge turnoff region stars, all highly magnified by microlensing, gives a mean
[Fe/H] of +0.29~dex.  This
implies that there there is an inconsistency between the Fe-metallicity
distribution of the microlensed bulge dwarfs  and  that derived by
the many previous estimates based on surveys of cool, luminous bulge giants,
which have mean [Fe/H] $\sim -0.1$~dex.
A number of possible mechanisms for producing this difference are discussed.
If one ascribes this inconsistency to systematic errors in the abundance analyses,
we provide statistical arguments suggesting that a substantial systematic
error in the Fe-metallicity for one or both of the two cases, 
bulge dwarfs vs bulge giants, is
required which is probably larger than can realistically be accommodated.

\end{abstract}


\section{Introduction \label{section_intro} }

High-magnification microlensing events present a
rare opportunity to obtain high resolution spectra
of otherwise  extremely faint dwarfs in the Galactic
bulge, which would require of order 100 hours of
observations on 8m class telescopes under ordinary
circumstances.  Microlensing is itself very rare,
with only a fraction $\tau \sim 10^{-6}$ of stars
being microlensed at any given time, even toward the
Galactic bulge where the density of lenses is
exceptionally high.  Events that are magnified
by a factor $A$ are rarer still by a factor $A^{-1}$.
And finally, the high-magnification
lasts only $A^{-1}t_{\rm E}$, where
$t_{\rm E}\sim 30\,$days is the Einstein timescale
of the event.  So there are formidable problems
predicting high-magnification episodes sufficiently
far in advance to arrange spectroscopic observations
from 8m class telescopes.

Nevertheless, two groups, Microlensing Observations in
Astrophysics (MOA) and the Optical Gravitational Lens
Experiment (OGLE) find a total of about 800 microlensing
events per year, of which the Microlensing Follow Up
Network\footnote{http://www.astronomy.ohio-state.edu/$\sim$microfun/} 
($\mu$FUN) is able to identify about 10 as
high-magnification events.  During the 2008 season, the additional
challenges posed by getting spectra on short notice
were overcome for three of these events, bringing
the total number of bulge dwarfs with high-magnification spectra
to seven.   There are four published analyses:
\johnstar\ \citep{johnson07},
\mystar\   \citep{cohen08},
\moa\      \citep{johnson08}, and
\bensbystar\ \citep{bensby09}.  
In addition, there is a spectrum
of OGLE-2007-BLG-514S taken by M. Rauch and G. Becker with
an as yet unpublished analysis by C.~Epstein \etal.

Here we analyze the two remaining high-mag
bulge-dwarf spectra from the 2008 season,
\moaten\ and \moaeleven, which, remarkably, peaked
on successive nights over Africa and were both observed
as they were falling from their peak at the beginning
of the Chilean night using the Magellan Clay telescope.  With
the addition of these two stars,
the sample microlensed bulge main sequence turnoff region
stars with high resolution, high quality spectra and
published detailed abundance
analysis becomes six stars; we refer to them 
collectively as the six microlensed dwarfs.

The ability to obtain high resolution, high quality spectra
of Galactic bulge stars and to carry out a detailed abundance
analysis offers an unbiased way to determine the metallicity
distribution of stars in the Galactic bulge, as well as their
detailed chemical inventory.  The goal of the present paper is
to carry out detailed abundance analyses for the two additional
microlensed bulge dwarfs (\S\ref{section_abunds}). Then 
in \S\ref{section_discussion} we use the 
six microlensed dwarf sample
to study the bulge metallicity distribution function
as well as their abundance ratios, and to
compare them to the results obtained by a number of surveys
of giants in the Galactic bulge.

\section{Observations}

\moaten\ and \moaeleven\ were observed on two consecutive nights in
July 2008 using the MIKE spectrograph \citep{bernstein03}
on the  6.5~m Magellan Clay Telescope
at the Las Campanas Observatory by I.~Thompson and G.~Burley.
Details of the exposures are given in Table~\ref{table_stars}. 
Spectroscopic exposures for \moaeleven\ began at UT 22:58 just after
sunset at airmass 1.93 (4.1 hours east of the meridian,
so the initially large airmass decreased quickly) when it was magnified
by a factor of 190; the star was just past its maximum brightness
of $I \sim  13.5$~mag and fading at that time.  The photometry
of this microlensing event is consistent with a point source
being magnified by a perfect point lens.

Spectroscopic exposures of \moaten\ began with MIKE the
following night at UT 22:51 at airmass 1.87 at the same hour angle
as for \moaeleven.  \moaten\ was brighter than \moaeleven\ at the time of
observation by $\sim$0.8~mag.  
A narrower slit 0.5~arcsec wide was used
to isolate \moaten\ from  a close companion roughly 2~mag fainter.
Fortunately the seeing that night was very good (0.6~arcsec) after the
first half hour (i.e. once the airmass became reasonable),  and
the companion rotated further away from the slit with  time. 
Thus, even with the narrower slit and consequently higher spectral resolution,
a high signal-to-noise ratio per spectral resolution element was
achieved for the spectrum of this star. 

The light curve of  \moaten\ shows pronounced finite-source effects,
with the lens exiting the limb of the source about 20 minutes
before the start of spectroscopic observations.  In addition,
the light curve shows much smaller deviations from standard
point-lens microlensing due to a companion to the lens
(J.~Janczak et al. 2009, in prep). 
Johnson \etal\ (2009, in preparation) has explored the impact of 
differential
amplification across the surface of a dwarf near the main sequence turnoff
as it affects an abundance analysis; she
finds it to be negligible compared to the uncertainties
in the abundances.

\section{Stellar Parameters \label{section_params} }

The microlensed bulge dwarfs suffer from substantial reddening
whose exact value is unknown.  We therefore rely purely
on their spectra to determine their stellar parameters.
The classical technique of excitation equilibrium for 
the set of the many Fe~I lines measured is used to
find \teff.  Then the microturbulent velocity $v_t$ is
set by requiring  the deduced Fe abundance to be independent
of the equivalent width \eqw\ for the same set of lines.  The surface gravity
is set by requiring ionization equilibrium 
between neutral and singly ionized Fe;
the ionization equilibrium for Ti in both of the
stars is then extremely good.
If the deduced [Fe/H] is substantially different from that
assumed to construct the model atmosphere, the process is repeated
with the [Fe/H] determined from the initial pass used for the model
atmospheres.
Throughout this process, we choose to ignore
lines with \eqw\ exceeding
130~m\AA\ due to the difficulty of properly including
their damping wings in the \eqw\ measurements.  
Features bluer than 5200~\AA\
were ignored unless the species had very few other detected lines
as the signal-to-noise ratio decreases rapidly at bluer wavelengths
due to the high reddening along the line of sight
to the Galactic bulge.

\begin{figure}
\epsscale{1.0}
\plotone{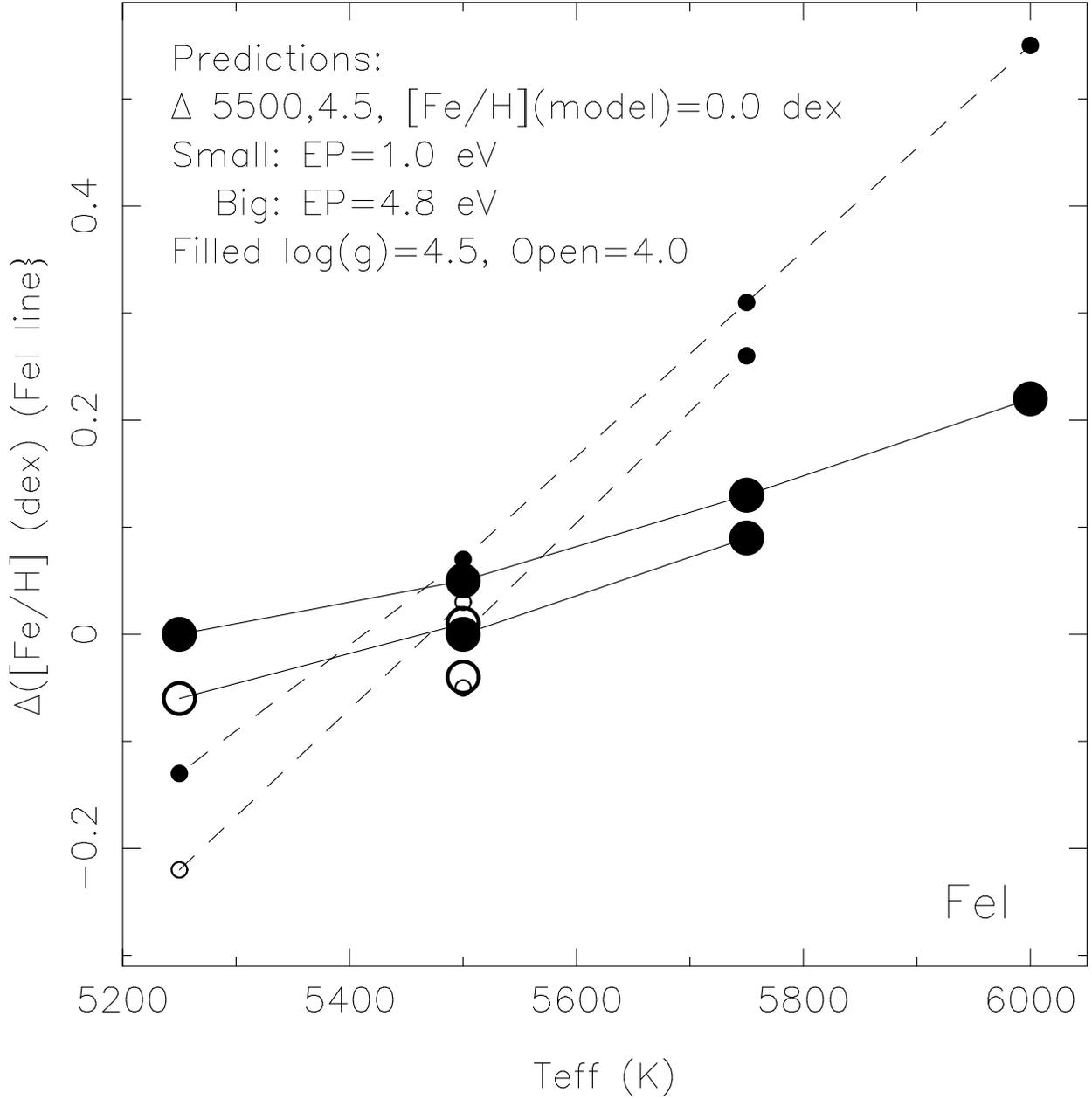}
\caption[]{Dependence of deduced abundance [Fe/H] from a weak
Fe~I line of fixed \eqw\ on \teff\ and \grav\ for lines of low 
($\chi = 1.0$~eV, small symbols connected by dashed lines)
and high ($\chi = 4.8$~eV, large symbols connected
by solid lines) excitation potential.  Open symbols
denote model atmospheres with \grav\ = 4.0~dex, filled symbols denote those
with \grav\ =4.0~dex.
The vertical axis is the difference in derived [Fe/H]
from the Fe~I line with respect to the
model with \teff\ = 5500~K, \grav\ = 4.5~dex, and [Fe/H] solar.
Increasing [Fe/H] of the model atmosphere by 0.5~dex increases the deduced
[Fe/H] by 0.05~dex.  Note the low sensitivity of high $\chi$ Fe~I
lines to \teff, \grav, and also to the adopted [Fe/H] for the model.
\label{figure_fe1_weakline}}
\end{figure}

Because the spectra are not perfect, and being concerned
about the convergence of this scheme onto the correct
set of stellar parameters,
we decided to develop a technique for determining [Fe/H],
at least approximately,
that might bypass some of these issues and indicate the magnitude
of some of the uncertainties in a more direct fashion.
Following in the spirit
of the line ratio method developed by \cite{gray91}
and used by \cite{biazzo07}, we looked for something
easy to measure
based purely on aspects of the spectrum that have a strong dependence
on metallicity, but little dependence on any other stellar parameter.
As a guide we constructed plots based on detailed abundance analyses
of the behavior of weak lines of species with many detected 
absorption lines as a function
of the set of adopted stellar parameters
\teff, \grav, [Fe/H] of the model atmosphere, and $v_t$ that would
enable us to isolate metallicity from them.
Figure~\ref{figure_fe1_weakline}
illustrates the best case we found for stars in the region
of the main sequence turnoff, namely [Fe/H]
derived from Fe~I absorption lines with
high excitation ($\chi > 4$~eV), which show low sensitivity
to changes in \teff\ of $\pm$250~K or of
\grav\ of $\pm$0.5~dex within the regime of interest.
Although not shown on the figure, we note that
increasing [Fe/H] of the model atmosphere by 0.5~dex increases the deduced
[Fe/H] by only 0.05~dex.  Using high excitation Fe~I lines, the 
final derived [Fe/H] from a detailed abundance
analysis is bound to be close to the true value even
if the adopted stellar parameters are slightly off. 
The weak dependence of the behavior of such lines
on \teff\ is a result of the competition between ionizing
Fe~I when \teff\ is increased versus increasing the population
in the high excitation state from which the absorption features
arise.  Fe~II lines  with $\chi \sim 0$~eV show a similar
behavior with \teff, but have much more sensitivity to changes
in \grav\ than do Fe~I lines.

The resulting stellar parameters for \moaten\ and \moaeleven\
are listed in Table~\ref{table_params}, which also gives
the slopes between deduced [Fe/H] abundances and the
excitation potential, \eqw, and $\lambda$ of the set of
Fe~I lines with \eqw\ $< 130$~m\AA.  We see that extremely good results
(i.e. almost flat relations with slopes very close to 0)
were obtained for the first two (primarily sensitive to 
\teff\ and to $v_t$ respectively).  The slope with
wavelength for \moaten\ is somewhat larger than ideal but the
correlation coefficient is low ($<0.15$), and the total change over the span
of 2600~\AA\ covered is only 0.07~dex.
The uncertainty in \teff\ is related to the first slope,
which decreases by $\sim$0.05~dev/eV when \teff\ is 
increased by 250~K in this regime.  Assuming a reasonable
sample of low excitation Fe~I lines, so that the range 
of the measured lines covers $\sim$4~eV, we set the 
uncertainty in \teff\ to 100~K.  The uncertainty in \grav\ then 
follows by considering the error resulting in ionization equilibrium
of Fe~I should \teff\ be off by 100~K, which has to be
compensated for by changing \grav.  There is an additional
smaller uncertainty in \grav\
arising from the uncertainty in the value of
[Fe/H](Fe~I) -- [Fe/H](Fe~II) itself. 
We find an uncertainty
in \grav\ of 0.2~dex in \grav\ is appropriate.

We note that a fit to the H$\alpha$ profile in \moaten,
the star with the higher SNR spectrum, indicates \teff\ $\sim 5500$~K,
120~K less than that derived from the Fe~I line analysis.
We also compare our derived values of \teff\ with those
that would be inferred from the photometry of the two
microlensed bulge dwarfs.
Light curves were obtained in two colors, $V$ and $I$,
by the $\mu$FUN collaboration for \moaten\
and for \moaeleven\ during the microlensing event as part
of an effort to detect planets.
The color of red clump stars\footnote{The dereddened red clump 
in the Galactic bulge
is assumed to have $I_0 = 14.32$~mag and $(V-I)_0 = 1.05$~mag.}
in the field around each of the microlensed stars is easily determined.
The comparison of instrumental $(V-I)$  and $I$ of the
red clump and the microlensed star then yield
$(V-I)_0$  and $I_0$ of the star, under the assumption
that it suffers the same extinction as the clump.
If the microlensed star is further assumed to lie
at the same distance as the clump, then 
the star's absolute magnitude $M_I$ can be calculated.
This yields
$(V-I)_0 = 0.70$~mag  for \moaten\ and $(V-I)_0 = 0.66$~mag for \moaeleven,
with $I_0 = 17.94$~mag for \moaten\ and 18.37~mag for \moa311;
$I$ is in the Cousins system.  The Sun has 
$V-I = 0.688\pm 0.014$~ mag \citep{holmberg06}, which would
suggest that \teff\ for these two microlensed stars is
quite close to that of the Sun.  Given the uncertainties
in the photometry and the probability of small spatial variations
in the reddening across the field,
this is in good agreement
with the \teff\ derived directly from the spectra of \moaten\
and of \moaeleven\ of Table~\ref{table_params}.  These independent
determinations of \teff, together with their uncertainties, 
are summarized in Table~\ref{table_teff}.

We consider whether the derived parameters are consistent with
the star being in the Galactic bulge by comparing \grav\
derived from the spectra with that derived from the photometry.
$I_0$ is converted into a total luminosity assuming a distance
to the Galactic center of 8~kpc.  Then using the derived
\teff, and assuming a mass of 1$M_{\odot}$, we predict
\grav(phot).  The agreement between \grav(phot) and
\grav(spec) is reasonable, with
differences of 0.1~dex \moaten, and 0.3~dex
for \moaeleven.  

\begin{figure}
\epsscale{0.85}
\plotone{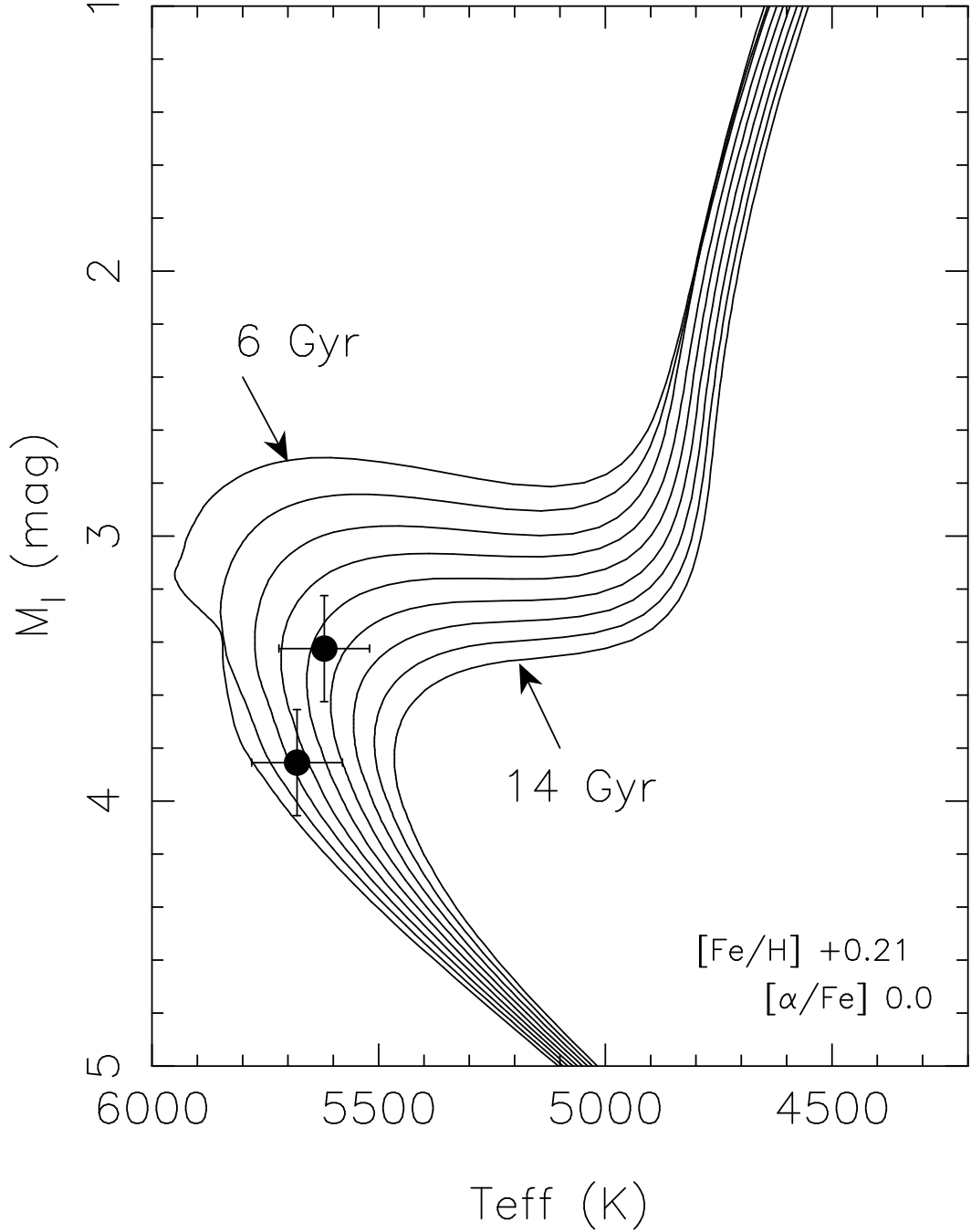}
\caption[]{A CMD with axes \teff\ and $M_I$ is shown with
the positions of the microlensed bulge dwarfs
\moaten\ and \moaeleven\ (the fainter of the two) as well as
with isochrones from the Dartmouth Stellar Evolution Database
\citep{dartmouth} with [Fe/H] +0.21~dex and [$\alpha$/Fe] = 0.0~dex.
The isochrones range in age from 6 to 14~Gyr in 1~Gyr increments.
\label{figure_age}}
\end{figure}

The ages of the microlensed bulge dwarfs are determined
by comparing $M_I$ as a function of \teff\ from
the relevant [Fe/H] and [$\alpha$/Fe] isochrones
of the grid of the Dartmouth Stellar Evolution Database
\citep{dartmouth}, as shown in Fig.~\ref{figure_age}.
The results for \moaten\
and \moaeleven\ are given in the last column
of Table~\ref{table_stars}; they are
consistent to within the errors with that of the Galactic
bulge population inferred from HST imaging, $\sim$10~Gyr,
by \cite{feltzing00}, see also \cite{zoccali03}.

\section{Abundance Analysis \label{section_abunds}}

Once the stellar parameters were determined, the abundance
analysis was carried out in a manner identical
to that for \mystar\ as described in \cite{cohen08}; in particular
it was a differential analysis with respect to the Sun.
In preparing the line list only features redder than 5200~\AA\
were used, unless there were none that red for a particular species,
due to increased crowding toward the blue
and to the high reddening.
Lines with \eqw\ $> 130$~m\AA\ were rejected unless the species
did not have at least a few suitably weak lines.  The exceptions
are the 5680~\AA\ Na doublet\footnote{The NaD lines are too corrupted by
interstellar features along the line of sight through the disk
to the bulge, and were ignored.} and the K~I resonance line
at 7700~\AA\ in both stars.  Also one Mg~I, one Si~I line,
and the only two Cu~I lines detected in \moaten, each
of which had \eqw\ $< 140$~m\AA, were retained.
The equivalent widths are given in Table~\ref{table_eqw}; their major uncertainty
results from the definition of the continuum level. 

We used a current version of the LTE
spectral synthesis program MOOG \citep{moog}.
We employ the grid of stellar atmospheres from \cite{kurucz93}
with [Fe/H] = 0.0 and +0.5~dex having solar
abundance ratios  
without convective overshoot \citep{no_over}  and with the most recent opacity
distribution functions. 
Non-LTE corrections were not 
included, as this is a differential analysis with 
respect to Sun, and the stellar parameters of both of these stars are fairly close
to those of the Sun.   Hyperfine structure corrections were used as
appropriate; see \cite{cohen08} for details.

The  deduced abundances for \moaten\ and for \moaeleven\  
are given in Tables~\ref{table_abunds_moa310} and \ref{table_abunds_moa311}
respectively, with
derived absolute abundances (second column), 
the abundances relative to the Sun (fifth column), and 
the abundance ratios [X/Fe] (seventh column). 
The abundance ratios use either 
Fe~I or Fe~II as the reference depending on the ionization
state and mean excitation potential of the measured lines of species
under consideration.  The 1$\sigma$ dispersion
around the mean for each species is given as $\sigma_{obs}$.
This is calculated from the set of differences between the
deduced solar abundance for the species in question and that 
found for a microlensed bulge dwarf
for each observed line of the species.  Thus
neither random nor systematic errors in the $gf$  values contribute
to $\sigma_{obs}$.

While the absolute abundance for a given species listed 
in Tables~\ref{table_abunds_moa310} and \ref{table_abunds_moa311}
will be affected
by any systematic error in the $gf$ values of the lines we
use here,  relative abundances [X/Fe] will not
since we have carried out a differential analysis with
respect to the Sun. An uncertainty
for [X/Fe] for each species, $\sigma_{pred}$, is calculated 
summing five terms  combined in quadrature representing
a change in \teff\ of 100~K, the corresponding uncertainty in \grav\
of 0.2~dex, a change in $v_t$ of 0.2~\kms, and 
a potential 0.25~dex  mismatch between
[Fe/H] of \mystar\ versus the value +0.5~dex of the model atmospheres
we are using.   (Tables~3 and 4 of \citealt{cohen08} give the
values of these four individual terms for each species.)
The fifth term, the contribution for errors in \eqw, is 
set to 0.05~dex if only one or two
lines were measured; for a larger number of
detected lines we adopt $\sigma_{obs}/\sqrt{N(lines)}$
for this term. 
This is added in quadrature to the other
four terms to determine our final uncertainty estimate
given in the final column of  
Table~\ref{table_abunds_moa310} and of Table~\ref{table_abunds_moa311}.
We note that the total uncertainty in [Fe/H] so derived is
0.09~dex when the large set of Fe~I lines is used.

\begin{figure}
\epsscale{0.85}
\plotone{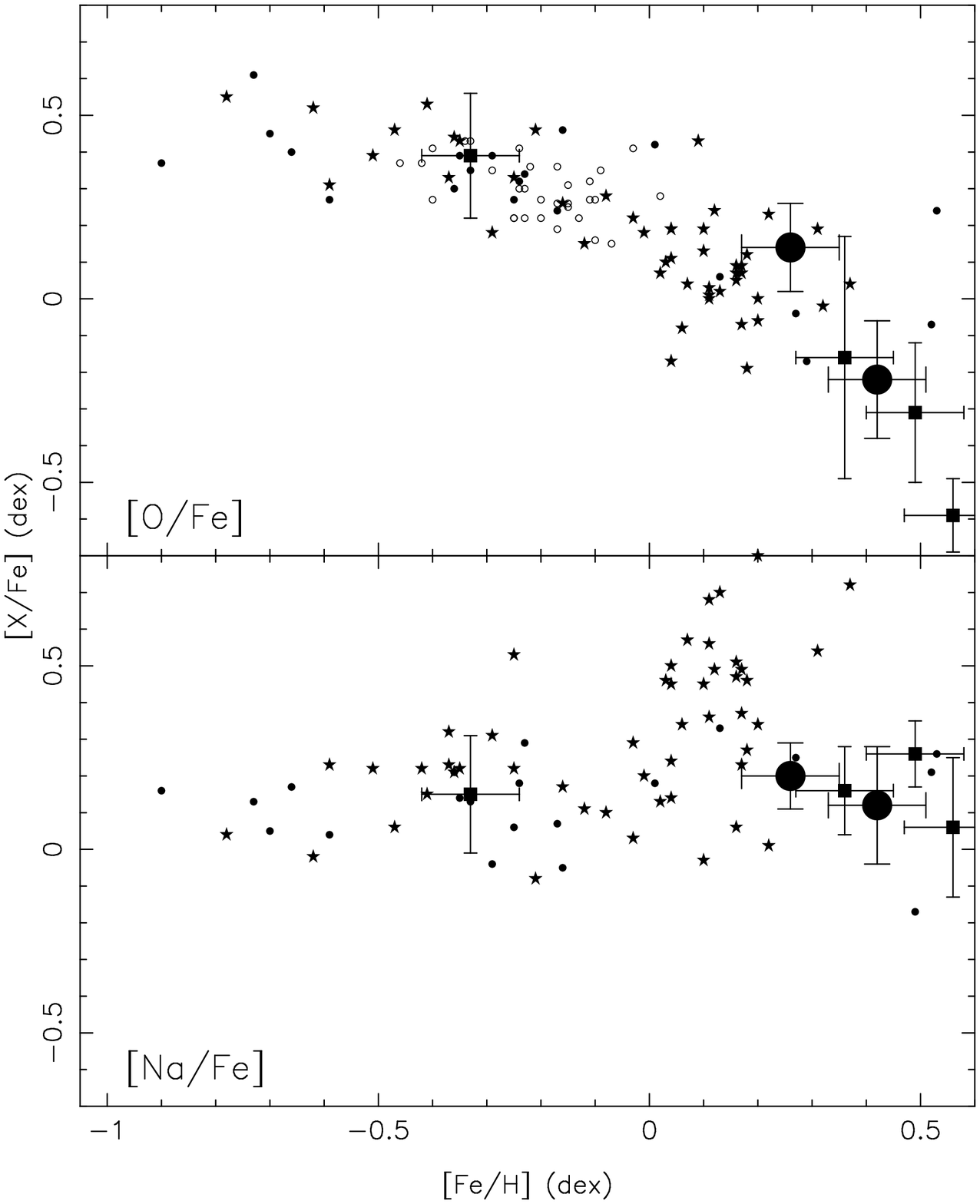}
\caption[]{Abundance ratios
[O/Fe] (upper panel) and [Na/Fe] (lower panel) are shown as a function
of [Fe/H]. 
\johnstar\ \citep{johnson07},
\mystar\   \citep{cohen08},
\moa\      \citep{johnson08},
\bensbystar\ \citep{bensby09}, and, from the present paper,
\moaten\ and \moaeleven\  are shown as large filled circles;
error bars are shown for them as well.
Samples
of bulge M and K giants of \cite{fulbright07} (small filled circles), 
\cite{rich05}  (small open circles),
\cite{lecureur07} (small stars),
and for M giants in the inner bulge from \cite{rich07} 
(small open circles) are also shown; their errors are somewhat smaller
than those of the microlensed dwarfs.
\label{figure_abund_ratios1}}
\end{figure}

\begin{figure}
\epsscale{0.9}
\plotone{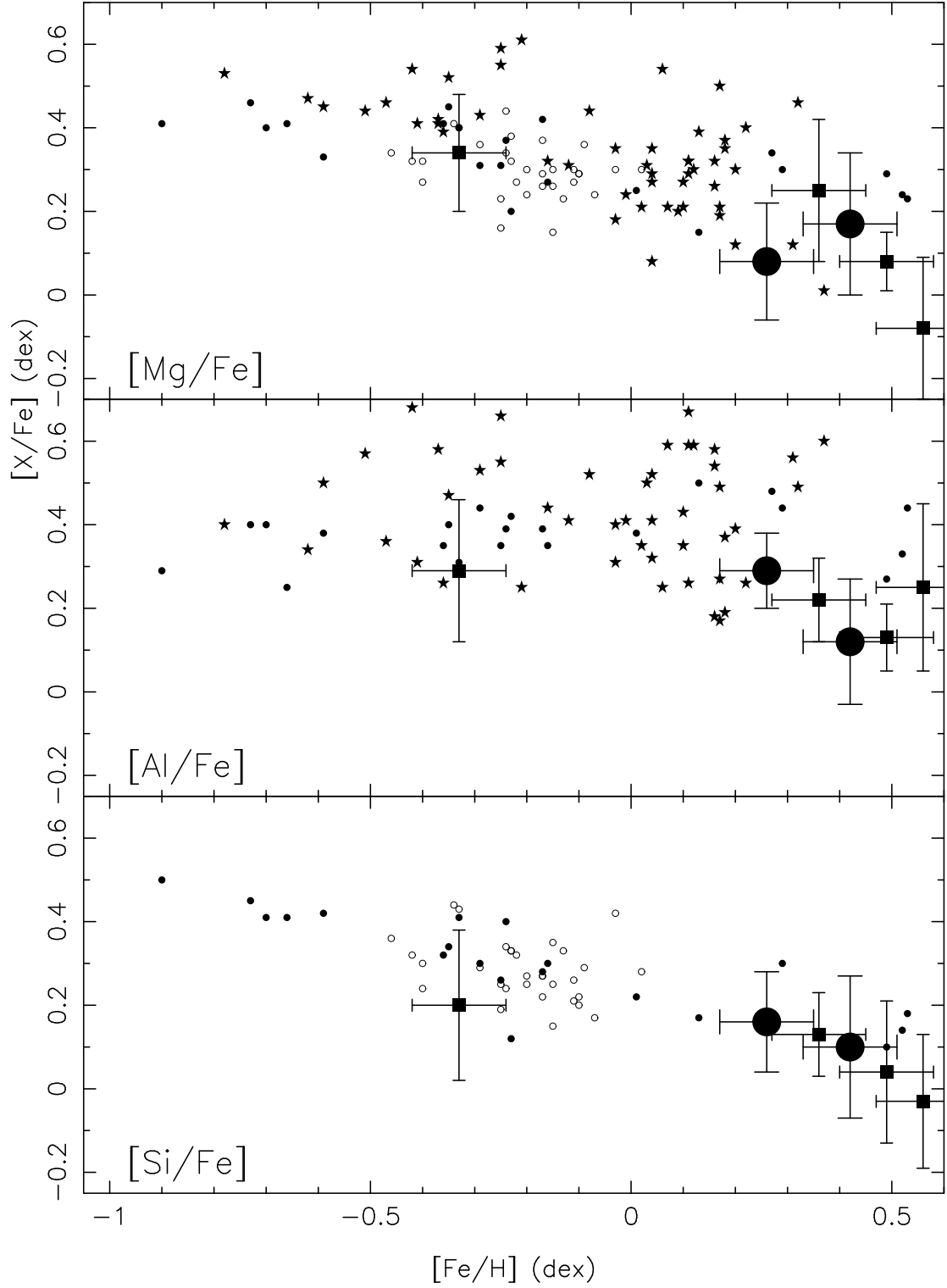}
\caption[]{The same as Figure~\ref{figure_abund_ratios1} for
[Mg/Fe] (upper panel), [Al/Fe] (middle panel) and [Si/Fe] (lower panel).  
The symbols are the same
as in Figure~\ref{figure_abund_ratios1}.
\label{figure_abund_ratios2}}
\end{figure}

\begin{figure}
\epsscale{1.0}
\plotone{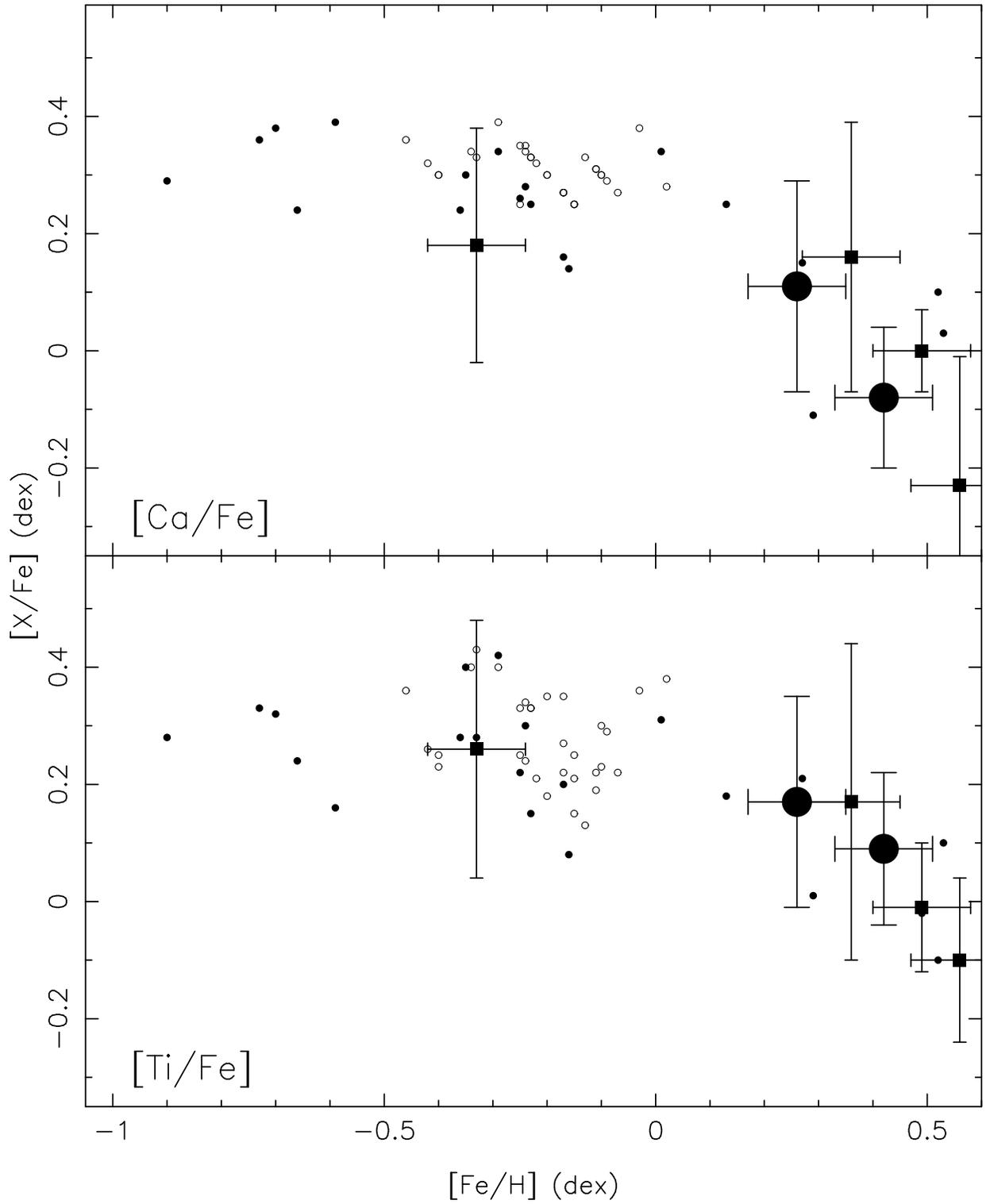}
\caption[]{The same as Figure~\ref{figure_abund_ratios1} for
[Ca/Fe] (upper panel) and for [Ti/Fe] (lower panel).  The symbols are the same
as in Figure~\ref{figure_abund_ratios1}.
\label{figure_abund_ratios3}}
\end{figure}

The key result of the abundance analysis is the high
Fe-metallicity found for the two microlensed Galactic bulge
stars in the region of the main sequence turnoff, [Fe/H] $= +0.41$~dex
for \moaten\ and +0.26~dex for \moaeleven.  The abundance
ratios are also of great interest. Figures~\ref{figure_abund_ratios1},
\ref{figure_abund_ratios2}, and \ref{figure_abund_ratios3} show
selected abundance ratios as a function of [Fe/H] 
for the six
microlensed bulge dwarfs.  These are compared
with abundance ratios from surveys of  Galactic
bulge giants by \cite{fulbright07}, 
\cite{rich05},
\cite{lecureur07}, and \cite{rich07}.
The figures demonstrate that to within the
 uncertainties microlensed bulge dwarfs have
abundance ratios [X/Fe] consistent with those of
Galactic bulge giants  at the same Fe-metallicities.
Comparisons between bulge giants and thick and thin disk
stars are given by \cite{fulbright07} and \cite{lecureur07}, while
detailed studies separating thick and thin disk stars, as
well as halo stars, via their abundance ratios and trends
with [Fe/H] include \cite{reddy03}
\cite{mashonkina04}, and \cite{bensby05}.

\section{The Metallicity Distribution of the 
Galactic Bulge \label{section_discussion} }

We have presented detailed abundance analyses for two additional
microlensed bulge main sequence turnoff region stars, so there
are now six that have been observed at high 
spectral resolution within the past three years
and for which detailed abundance analyses have been completed;
the references for the additional four are given in \S\ref{section_intro}.
Only one of the six  falls below solar
metallicity, at [Fe/H] $= -0.32$~dex; all the others are well
above solar, with the mean of the six being $\langle \rm{[Fe/H]}\rangle = +0.29$~dex.

\begin{figure}
\epsscale{1.0}
\plotone{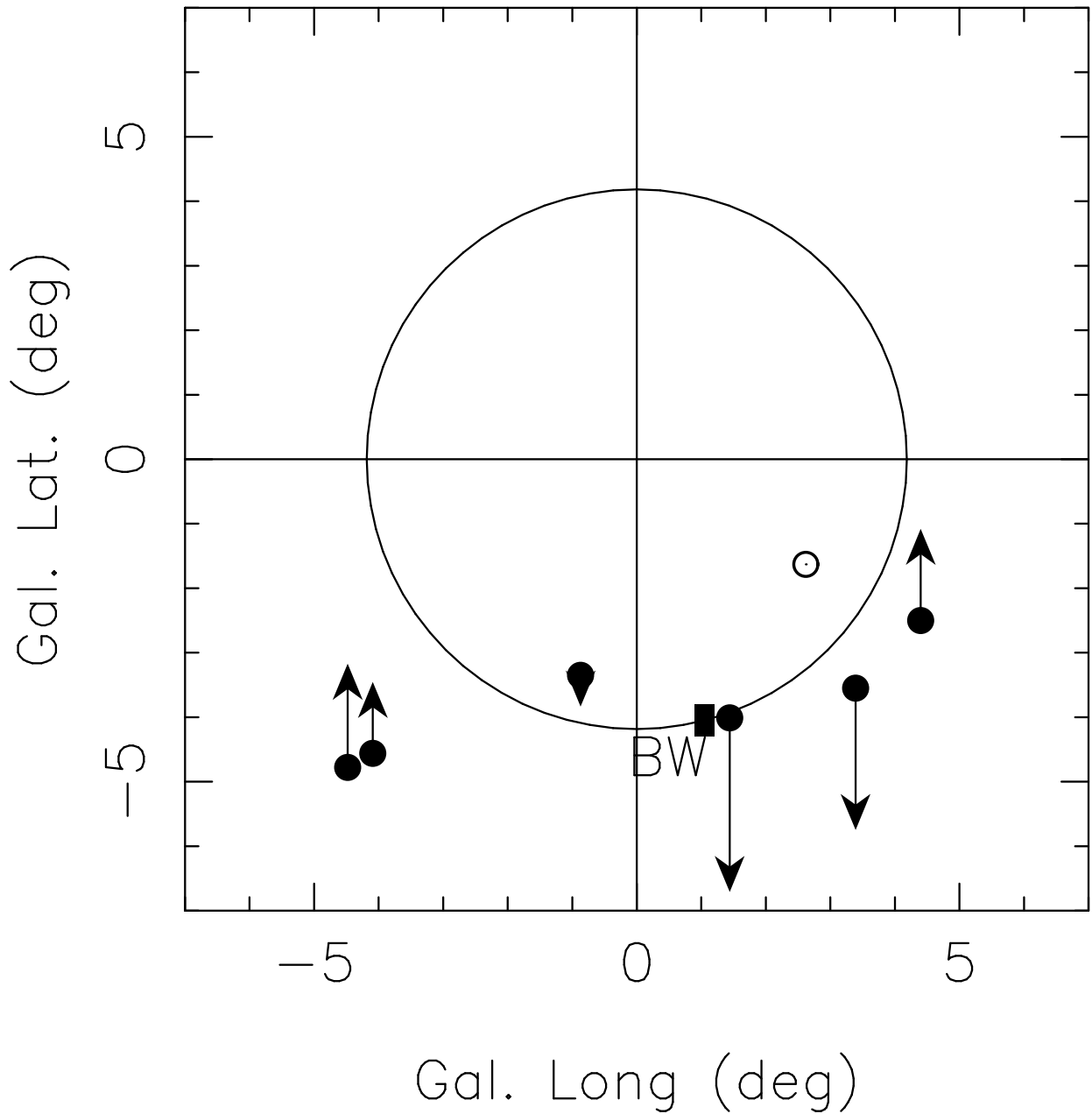}
\caption[]{The distribution in Galactic latitude and longitude of
the six microlensed bulge stars. 
The  heliocentric radial velocity for
each star is indicated by an arrow, upward being positive, with a
scale of 70 \kms\ per degree.
The small open circle denotes the unpublished spectrum of
OGLE--2007--BLG--514 taken by M.~Rauch being analyzed by C.~Epstein. 
Baade's Window is marked by the
filled rectangle, and its Galactocentric radius is indicated by a circle.
\label{figure_onsky}}
\end{figure}

The positions on the sky of the six microlensed bulge dwarfs
are shown in Figure~\ref{figure_onsky}. (The
magnitude of their
radial velocities are also indicated on this figure.)
The location of 
Baade's Window is marked.  This
is the field closest to the center of the Milky Way
with reddening low enough
that its bulge giants can be studied in detail in the optical with 
the current generation of large telescopes, and has
thus been the subject of many recent surveys
at the VLT \citep[see, e.g.][]{zoccali08} and at Keck
\citep[][among others]{fulbright06}.  The circle 
marks the region with projected Galactocentric distance equal to that
of Baade's Window.  It is important to note that five of the six
microlensed bulge dwarfs are slightly outside that circle;
only one is slightly within it.  This means that the population
we are sampling via microlensing should be essentially 
identical to the population 
sampled by studies of the giants in Baade's Window.
\cite{zoccali08} detected a small radial gradient in the
mean metallicity with Galactocentric distance of
0.6~dex/kpc (0.08~dex/deg on the sky at the distance of the Galactic center).
The gradient was established between Baade's Window and bulge fields
with larger projected $R_{GC}$.  
This would suggest  that the mean Fe-metallicity of
the sample of microlensed bulge dwarfs should be $\sim$0.05~dex
lower than that of Baade's Window.

\cite{zoccali08} recently redetermined the Fe-metallicity distribution
function in Baade's Window using a
sample of 204 luminous K giants with $14.2 < I < 14.7$~mag
spanning a wide range in $V-I$ color ($1.53 < (V-I) < 2.62$~mag)
so as to cover the full metallicity range within the stellar population
of the Galactic bulge.  There is also a sample of 
$\sim$200 red clump
giants in the Baade's Window discussed in \cite{lecureur07}.
Red clump stars are biased against low metallicities,
where RR~Lyrae and blue horizontal branch stars would be expected
instead, but
should not be biased at the [Fe/H] values relevant here, [Fe/H] $> -1$~dex.
\cite{zoccali08} combine the two datasets for a total sample of $\sim$400 giants
in this field. 

\begin{figure}
\epsscale{1.0}
\plotone{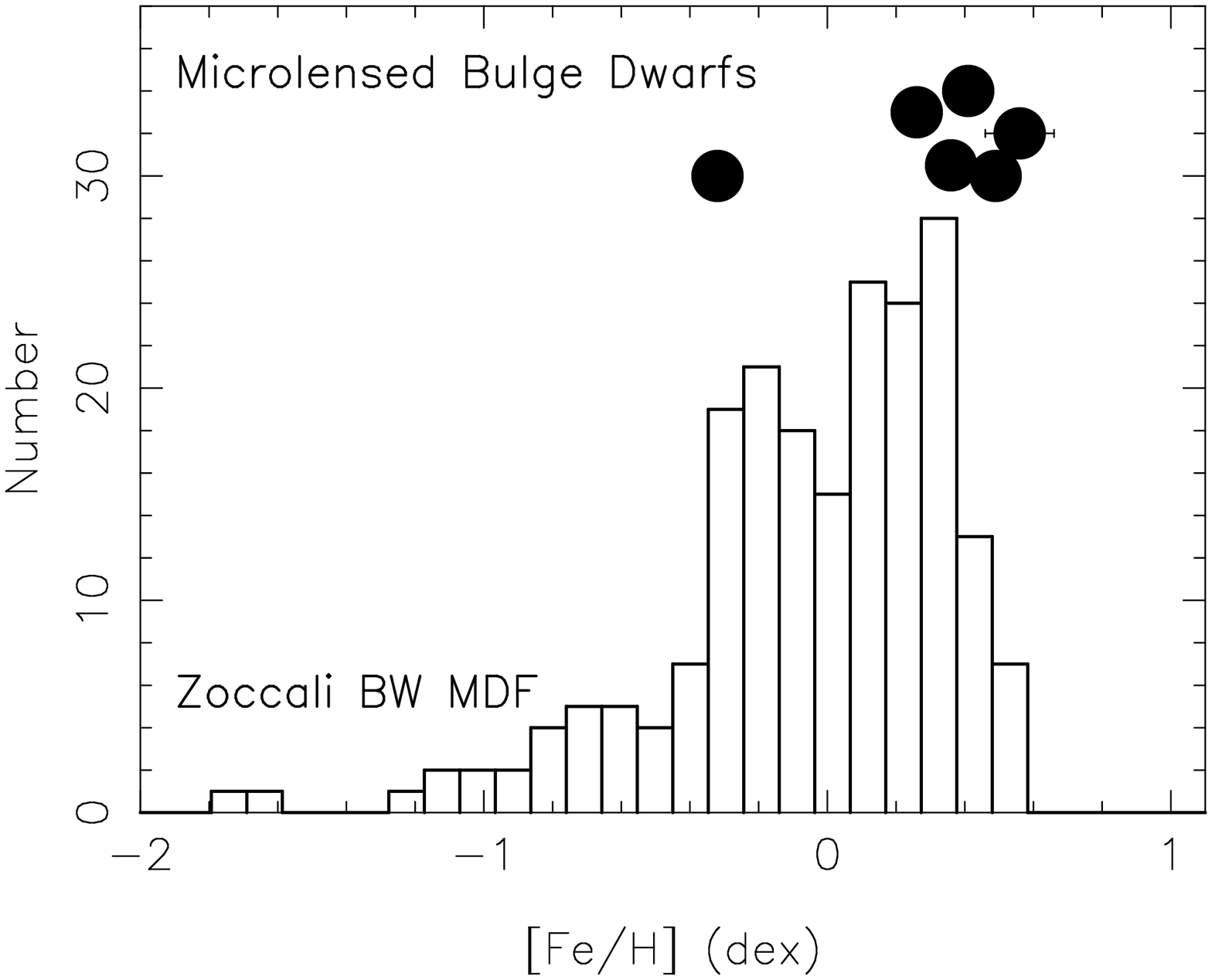}
\caption[]{The Fe-metallicity distribution from Zoccali et al (2008)
for stars in Baade's Window is shown.  The 6 microlensed dwarfs
with high resolution spectra and detailed abundance analyses, including
the two published here, are shown as filled circles: see 
\cite{cohen08} for \mystar, \cite{johnson07} for \johnstar,
\cite{johnson08} for \moa, and \cite{bensby09} for \bensbystar\
for the other four stars.  A typical uncertainty
in [Fe/H] for the microlensed bulge dwarfs is shown for the most metal-rich
star.
\label{figure_feh_hist}}
\end{figure}

Figure~\ref{figure_feh_hist} compares the Fe-metallicity distribution
function  recently determined by \cite{zoccali08}
in Baade's Window with that of 
the six microlensed bulge dwarfs.   The distributions are clearly
different.  The microlensed bulge dwarfs reveal a significantly
higher mean Fe-metallicity than do the giants studied by \cite{zoccali08},
who find a mean [Fe/H] of $-0.04$~dex for the 204 K giants
and $+0.03$~dex for the red clump stars from \cite{lecureur07}.
This is considerably lower than that of the six microlensed bulge dwarfs.
Furthermore, K and M giants closer to the Galactic center than Baade's Window
at ($l,b)~ = ~ (0^{\circ},-1^{\circ})$
have been probed through high resolution infrared spectroscopy
by \cite{rich07}, who also find a low mean Fe-metallicity, $-0.22$~dex,
and no sign of a radial gradient in metallicity for
$R_{GC}$ inward from Baade's Window \citep[see also][]{cunha06}.

To evaluate the statistical significance of this difference in
Fe-metallicity, we drew six stars at random from the sample of
\cite{zoccali08}, eliminating stars in the globular cluster NGC~6822, which
is in Baade's Window, and also those with highly uncertain
[Fe/H] as indicated by the quality codes in their table.
We took the  average [Fe/H], which we call the six star mean.
Note that the microlensed bulge star with by far the lowest [Fe/H] is
actually a subgiant; it is the only subgiant among the six and
is quite discrepant in [Fe/H]
from the other five microlensed bulge dwarfs. 
The results for 40,000 such trials are given
in Table~\ref{table_prob} as the percentage of trials
where the mean [Fe/H] for six stars drawn from the bulge giant sample
equaled or exceeded that of the set  of six microlensed bulge dwarfs.

If the [Fe/H] values of the large sample of bulge giants from
\cite{zoccali08} are correct, and those of the six microlensed
dwarfs are correct as well, then the probability that the two metallicity
distribution functions are identical 
is very small, $\sim 4 \times 10^{-3}$, ignoring any radial gradient, which
would further reduce the tabulated probabilities for metallicity increasing
as Galactocentric radius decreases. A K-S test also indicates
a very low probability that the two metallicity distributions
are the same, 1.9\%. However, if there are 
systematic errors in the
metallicity scale of either (or of both), and they act in the right direction,
the probability of this happening by chance increases.  Therefore
Table~\ref{table_prob} also contains the probability in the case
of  systematic
offsets of the correct sign ranging from 0.05 to 0.20~dex in size.
A systematic difference in Fe-metallicity scale 
between the two samples of 0.20~dex such that either the
bulge giant metallicities are underestimated or those of the microlensed
dwarfs are overestimated is
required before the probability reaches 20\%. 

\cite{zoccali08} quotes a ``conservative'' uncertainty in [Fe/H] of
an individual giant as
$\pm 0.2$~dex, including possible systematic errors.  A substantial systematic
error in [Fe/H] is required to produce consistency
between the microlensed bulge stars and the K (and M) giant samples.
If only \teff\
is changed and one looks at the Fe-metallicity derived from Fe~I lines
(which is only logical, since there are far fewer Fe~II lines
detectable), a 0.2~dex change corresponds to a 400~K systematic error for
the microlensed stars near the main sequence turnoff,
as was shown in Figure~\ref{figure_fe1_weakline}, and to 
at least a 500~K systematic error if the problem lies in the cool
giants, as at such low temperatures, iron is almost entirely
neutral.

This level of systematic error is  
larger than the uncertainty in the absolute Fe-metallicity 
of the microlensed
dwarfs, as their spectra can be compared directly to the 
solar spectrum.  
\cite{bensby09}  includes a comparison of [Fe/H] for
the previously published  four microlensed bulge dwarfs 
derived independently, with different
codes, different grids of model atmospheres, and different schemes
for determining the stellar parameters, by T.~Bensby, J.~Cohen, and
J.~A.~Johnson; the agreement among the analyses by the three independent groups
is quite good,  $\pm 0.06$~dex.  
On the other hand,  the analysis of the cool giants and the
determination of their stellar parameters is much
more difficult.  However, the required error in \teff\ for
the bulge giants is even larger, and seems very unlikely.
Furthermore many independent
groups have surveyed giants in the Galactic bulge,
with similar results as to the mean Fe-metallicity,
so there is no  reason to assign the required systematic
error to them.

There are a number of consistency checks that have been or could
be carried out to test the validity of the absolute Fe abundances
between the bulge giants and the microlensed dwarfs. 
The scale of the Fe transition probabilities is not relevant for the dwarfs,
as a differential solar analysis was used.  But it is for the giants;
\cite{fulbright06} used a differential analysis with respect to the well
studied giant Arcturus, while \cite{lecureur07} use the spectrum of
the metal-rich
giant $\mu$~Leo to derive pseudo-$gf$ values appropriate for their
method of measuring \eqw\ and their grid of model atmospheres, while
their absolute scale for [Fe/H] is set by taking 
[Fe/H] for this star as +0.30~dex.  Checks of the determination
of the continuum level in the giant spectra, where this is quite difficult,
could be carried out with very high quality spectra of a few bulge giants.
Differences between the model atmosphere grids are probably not the
cause as several independent groups have participated both for
the dwarfs and for the giants.  However, 
systematic problems affecting all the chosen model atmosphere grids as \teff\
decreases such as overionization of Fe
could be contributing.  Arguments that studies
of members of a single open or globular cluster at a wide
range of luminosities
 show no such effect (see e.g. Santos et al 2009 vs.
 Boesgaard, Jensen \& Deliyannis 2009)
 re often not relevant to the present case 
 when examined in detail.
 For example, \cite{pasquini04} studied giants and dwarfs
 in the open cluster IC~1651 with [Fe/H] +0.10~dex.  However,
 their coolest and most luminous giant is 
 several hundred K hotter in \teff\ and 0.4~dex higher in \grav\
 than the hottest and least luminous of the bulge giants in
 the sample of \cite{fulbright06}.  Stars near the RGB tip
 are very rare, and are unlikely to be found in any open cluster,
 while no globular cluster with [Fe/H] $> +0.1$~dex is known
 in the Galaxy.  Furthermore the best abundances for the most
 metal-rich  clusters come from dropping in luminosity to the RHB, where
 \teff\ is considerably higher, and avoiding the RGB tip giants
 completely \citep[see, e.g.][]{cohen99}.

There is thus a clear discrepancy between the metallicity distribution
function in the Galactic bulge as sampled by microlensed main
sequence turnoff region stars and by luminous K and M giants.
While still more microlensed dwarfs with detailed abundance analyses
are highly desired to improve the statistics, we assume 
that this
difference is real and is not the result of systematic errors
producing suitable offsets in [Fe/H] derived from
the abundance analyses.

In our earlier paper \citet{cohen08} we offered the suggestion
that the highest metallicity giants have such high mass loss rates
that they do not get to the RGB tip before losing their entire envelope.
Possible evidence against this hypothesis is presented by 
\cite{zoccali08} on the basis of the luminosity function along
the RGB; \cite{clarkson08} comment that the metallicity distribution
of the bulge giants and of main sequence stars inferred from ACS/HST phometry
are consistent with each other.
An additional possibility is that we are sampling
a  ``young'' and metal-rich stellar population such as that found within
the inner 40~pc, where rather surprisingly massive young clusters exist
\citep{figer02}, presumably fed, at least in part, by mass loss from
bulge giants.  This runs into the problem that the corresponding 
high luminosity stars from such a population are not present 
in Baade's Window, as reinforced by the very recent ACS/HST study
of the Galactic bulge by \cite{clarkson08}.

A similar argument applies for any proposed special component
of the central region of our Galaxy such as an extension of the disk.
\cite{luck06} determined the metallicity gradient
for the Galactic disk 
from analysis of a large sample of Cepheid variables 
outside 4~kpc from the center to be $-0.06$~dex kpc$^{-1}$.
It is interesting to note that their deduced [Fe/H] reached
+0.3~dex at $R_{GC} = 4$~kpc, and if their linear fit is
extrapolated inward, would reach +0.5~dex at $R_{GC} = 1$~kpc.
A similarly metal-rich population of solar neighborhood disk stars
whose highly eccentric orbits have pericentric distances
as small as 3~kpc was identified by
\cite{grenon99} and \cite{pompeia02}.  
These super metal-rich old dwarfs  have [Fe/H] reaching up
to +0.4~dex and mean distance from the Galactic plane of only 220~pc.
But a rather puffed up disk would be required to contribute
significantly at Baade's Window, which is at $b \sim -4^{\circ}$ (560~pc).
Certainly over a very large range in $R_{GC}$ the 
vertical scale height of the thin disk is smaller than that.

Another possibility is that
the disk and/or halo contamination in the giant samples in Baade's Window
is larger than that calculated from Galactic models
by \cite{zoccali08} and others. Little is known of the detailed
structure of the disk and bulge in the region of the Galactic center.
Although disk and halo contamination of the giant samples are believed
to be small based on calculations using models of the stellar population
of the Galaxy \citep[see, e.g.][]{zoccali08},
the uncertainty in such corrections might be large.  The extensive
proper motion  studies in  bulge fields \citep[see, e.g. ][]{clarkson08} give
good determinations of the foreground disk contamination, but cannot easily address
the possible presence of the disk within the bulge itself provided  it
makes a minor contribution to the total stellar population in the bulge.
Disk contamination in the microlensed sample, which is a sample
of background sources, must be smaller than that
of the in situ giant samples, which probe the long line of sight to
the center; the probability of microlensing for a foreground disk star 
is much smaller
than for a star in the bulge itself, hence very biased strongly against
disk stars.

\section{Summary \label{section_summary}  }

We present detailed abundance analyses based on high dispersion and high
signal-to-noise ratio MIKE spectra taken with the  6.5~m Magellan Clay
Telescope of two highly microlensed Galactic
bulge stars in the region of the main sequence turnoff. 
Our stellar parameters were derived ignoring the available photometry out of
concern for the high and uncertain reddening toward the bulge, and rely
only on the spectra themselves.  They are based on the
classical criteria of Fe excitation equilibrium, and the ionization
equilibrium of Fe and of Ti, and are consistent to within the
adopted errors with that inferred from the H$\alpha$ profile
for the star with the higher quality spectrum, \moaten. 
We deduce \teff\ near 5650~K for both of
these stars. \moaten\ and \moaeleven\ appear to be at the distance
of the bulge with age $\sim$9~Gyr.

We suggest that the use of high excitation ($\chi > 4$~eV) Fe~I lines is the measure
of metallicity
most independent of the exact choice of values for stellar parameters
for such stars among the various possibilities we explored.  We note that
the available $V,I$ photometry for the two stars supports our choice of
\teff\ for each to within the photometric errors and the 
uncertainty of the reddening determination, which is
based on red clump stars in the bulge in the field around each of the microlensed
dwarfs.

We carry out a detailed classical abundance analysis using 1D stellar model
atmospheres and ignoring non-LTE.  Since this is done differentially
to the Sun and the two stars both have \teff\ within 160~K of that
of the Sun and \grav\ within 0.3~dex of the Sun, these choices seem
appropriate.
We find that \moaten\ has [Fe/H] = +0.41$\pm$0.09~dex and \moaeleven\ has +0.26$\pm$0.09~dex.
The abundance ratios for the $\sim$20 elements for which features could
be detected in the spectra of each of the two stars follow the trends
with [Fe/H] found among samples of Galactic bulge giants.  

Combining
these two bulge stars with the results from previous abundance analysis of
four other
Galactic bulge dwarfs, all highly magnified by microlensing, gives a mean
[Fe/H] of +0.29~dex for the six microlensed dwarfs, 
which rises to +0.41 when the lowest metallicity dwarf, which is actually
a subgiant with [Fe/H] very discrepant from the other five stars,
is removed.  On the other hand, the many large surveys of the metallicity distribution
function in the Galactic bulge carried out at the VLT \citep{lecureur07,zoccali08} and at 
Keck \citep[][among others]{fulbright06,rich07} from samples of cool, luminous bulge giants
give mean [Fe/H] $\sim -0.1$~dex.  This
implies that there is an inconsistency between the Fe-metallicity
distribution of the microlensed bulge dwarfs  and  that derived by
the bulge giants.  This difference is highly statistically significant
assuming that both the abundance analyses of the giant samples
and of the six microlensed dwarfs have been carried out correctly.

We provide statistical arguments suggesting that 
to produce consistency a substantial systematic
error in the absolute metallicity of Fe in one or both of the two cases,
 bulge dwarfs vs bulge giants, is necessary.  The required
offset which must act to either underestimate the metallicities for the giants
 or overestimate those of the microlensed dwarfs, or both of these,
 is  0.2~dex in [Fe/H], ignoring a radial gradient, which would only
increase this value.  Were a systematic offset of this size present,
the probability of the observed metallicity distribution functions
for these two groups of bulge stars in very different evolutionary
phases to be identical would rise to  15\%.

Since the microlensed main sequence region stars are usually analyzed
differentially with respect to the Sun, to which they are fairly close
in stellar parameters, the resulting systematic errors should be small.
Furthermore there are now
 multiple independent analyses for several of the microlensed
dwarfs \citep[see, e.g.][]{bensby09},
and there are several major independent surveys of bulge giants, suggesting
that it is unlikely that either the dwarfs or the giants or both
have major
systematic errors in 
their [Fe/H] determinations.
The contamination by foreground disk stars is predicted to be small 
for the giant samples; samples of bulge dwarfs selected through 
microlensing
should contain a considerably smaller fraction of
foreground disk stars.

A number of mechanisms for producing this difference are discussed, but none
seems compelling.  We clearly need a still larger sample of microlensed bulge
dwarfs to refine the systematic offset required to
achieve statistically identical Fe-metallicity distributions
and to eliminate completely the possibility that  a systematic
error of the required size may have occurred
in one or both of the Fe-metallicities 
between the bulge giants
and the bulge microlensed
dwarfs  before indulging in further speculations
 of the cause of this
discrepancy.  The rising interest in time-domain phenomena
has led to increased attention on how to handle these phenomena
efficiently at large telescopes, increasing sensitivity for
the handling of targets of opportunity.  In the past three years,
high dispersion spectra for six highly
microlensed bulge  dwarfs have been obtained at the Las Campanas or the
Keck Observatory.  With high hopes that the same will hold for the next
three years, we eagerly await future larger samples of microlensed bulge
turnoff region stars.

\acknowledgements

J.G.C. and W.H. are grateful to NSF grant AST-0507219 to JGC for partial
support. I.B.T. is grateful for support
NSF grant AST-0507325. A.G. was supported by NSF grant 0757888.
T. Sumi is grateful for a Grant-in-Aid for Young Scientists (B) and
Grant-in-Aid for Scientific Research on Priority Areas, ``Development  
of Extra-solar Planetary Science'' by the Ministry of Education, Culture,  
Sports, and Technology (MEXT) of Japan.
I.Bond is grateful to support from the Marsden Fund of the Royal Society of
New Zealand.

{}

\clearpage

\begin{deluxetable}{l rrrrr  r}
\tabletypesize{\footnotesize}
\tablenum{1}
\tablewidth{0pt}
\tablecaption{Properties of \moaten\ and \moaeleven
\label{table_stars}}
\tablehead{
\colhead{ID} & \colhead{Date of Obs.} & \colhead{Exp. Time} &
\colhead{Spec. Res} & \colhead{SNR\tablenotemark{a}} &
\colhead{$v_r$\tablenotemark{b}} & \colhead{Age\tablenotemark{c}} \\
\colhead{} & \colhead{} & \colhead{(sec.)} & \colhead{} &
\colhead{} & \colhead{(\kms)} & \colhead{(Gyr)} 
}
\startdata
\moaten & 8/7/2008 & 4x1800 & 41,000 & 115 & +77.5 & 9.5$\pm$2.0 \\
\moaeleven & 7/7/2008 & 4x1800 & 29,000 & 104 & $-$34.1 & 7.8$\pm$2.5 \\ 
\enddata
\tablenotetext{a}{Signal-to-noise ratio per spectral resolution element
in continuum at 6025~\AA\ (at the center of an echelle order).}
\tablenotetext{b}{Heliocentric radial velocity.}
\tablenotetext{c}{We use isochrones from the Dartmouth Stellar Evolution Database of
\cite{dartmouth}, see Fig.~\ref{figure_age}.}
\end{deluxetable}

\begin{deluxetable}{l rrrr rrr}
\tabletypesize{\footnotesize}
\tablenum{2}
\tablewidth{0pt}
\tablecaption{Stellar Parameters of \moaten\ and \moaeleven
\label{table_params}}
\tablehead{
\colhead{ID} & \colhead{$T_{eff}$} & \colhead{log($g$)} &
\colhead{[Fe/H]} & \colhead{$v_t$} &
\colhead{$\Delta$[X/Fe]/$\Delta$(EP)\tablenotemark{a}} &
\colhead{$\Delta$[X/Fe]/${\Delta}[W_{\lambda}/\lambda]$} &
\colhead{$\Delta$[X/Fe]/${\Delta}\lambda$} \\
\colhead{} & \colhead{(K)} & \colhead{(dex)} & \colhead{(dex)} &
\colhead{(\kms)} &
\colhead{(dex/eV)} & \colhead{(dex)} & 
\colhead{($10^{-4}$~dex/$\AA$)}
}
\startdata
\moaten    & 5620 & 4.3 & +0.5 & 1.0 & 0.013 & $-$0.055 & 0.256 \\
\moaeleven & 5680 & 4.1 & +0.3 & 1.2 & 0.006 & 0.054 & $-0.066$ \\ 
\enddata
\tablenotetext{a}{Typical range of EP is 4 eV.
This slope decreases by $\sim$0.05~dex/eV for an  increase
in \teff\ of 250~K.}
\end{deluxetable}

\begin{deluxetable}{l rrr}
\tablenum{3}
\tablewidth{0pt}
\tablecaption{Determinations of \teff\ Using Various Methods
For \moaten\ and \moaeleven
\label{table_teff}}
\tablehead{
\colhead{ID} & \colhead{} & \colhead{$T_{eff}$~(K)} \\
\colhead{} &  \colhead{Fe~I Lines} &  \colhead{$(V-I)_0$} &
      \colhead{H$\alpha$ Profile}
}
\startdata
\moaten    & 5620 $\pm$100 & 5800 $\pm$225\tablenotemark{a} & 5500 $\pm$150 \\
\moaeleven & 5680 $\pm$100 & 5640 $\pm$225\tablenotemark{a} & \nodata \\ 
\enddata
\tablenotetext{a}{We assume an uncertainty of 0.05~mag in $(V-I)_0$ 
due to uncertainty in the color of the red clump
and possible differential reddening between the clump and
these particular stars.}
\end{deluxetable}

\clearpage

\begin{deluxetable}{l  crrr  rr }
\tablenum{4}
\tablewidth{0pt}
\tablecaption{$W_{\lambda}$ for the Sample EMP Stars From the HES \label{table_eqw}}
\tablehead{
\colhead{$\lambda$} & \colhead{Species} & \colhead{EP} &
\colhead{log($gf$)} &
\colhead{\moaten} & \colhead{\moaeleven} \\
\colhead{($\AA$)} & \colhead{} &
\colhead{(eV)}    &   \colhead{}  & \colhead{(m$\AA$)} & \colhead{(m$\AA$)} 
}
\startdata
 6300.30 & O(OH) &   0.00 &  $-$9.780 &     7.1 &    44.7     \\
 7771.94 & O(OH) &   9.15 &   0.369 &    75.4 &    88.2     \\
 7774.17 & O(OH) &   9.15 &   0.223 &    68.2 &    83.4     \\
 7775.39 & O(OH) &   9.15 &   0.001 &    54.0 &    63.3     \\
 5682.63 & Na~I  &   2.10 &  $-$0.700 &   161.0 &   159.4     \\
 5688.19 & Na~I  &   2.10 &  $-$0.420 &   167.7 &   163.8     \\
 6154.23 & Na~I  &   2.10 &  $-$1.530 &    76.5 &    54.1     \\
 6160.75 & Na~I  &   2.00 &  $-$1.230 &    87.6 &    85.6     \\
 5711.09 & Mg~I  &   4.34 &  $-$1.670 &   135.0 &   121.4     \\
 6318.72 & Mg~I  &   5.11 &  $-$2.100 &    80.3 &  \nodata     \\
 6319.24 & Mg~I  &   5.11 &  $-$2.320 &    60.0 &  \nodata     \\
 6696.02 & Al~I  &   3.14 &  $-$1.340 &    77.9 &    73.2     \\
 6698.67 & Al~I  &   3.14 &  $-$1.640 &    46.0 &    32.3     \\
 5421.18 & Si~I  &   5.62 &  $-$1.430 &  \nodata &    85.3     \\
 5665.55 & Si~I  &   4.92 &  $-$2.040 &    74.2 &    63.9     \\
 5690.43 & Si~I  &   4.93 &  $-$1.870 &    66.8 &    74.1     \\
 5701.10 & Si~I  &   4.93 &  $-$2.050 &    59.4 &    53.0     \\
 5772.15 & Si~I  &   5.08 &  $-$1.750 &    81.4 &    90.0     \\
 5793.07 & Si~I  &   4.93 &  $-$2.060 &    70.6 &    66.0     \\
 5948.54 & Si~I  &   5.08 &  $-$1.230 &   117.3 &   113.6     \\
 6145.02 & Si~I  &   5.61 &  $-$1.440 &    61.3 &    59.1     \\
 6155.13 & Si~I  &   5.62 &  $-$0.760 &   132.5 &    99.9     \\
 6237.32 & Si~I  &   5.62 &  $-$1.010 &   100.3 &    98.0     \\
 6721.84 & Si~I  &   5.86 &  $-$0.939 &    78.3 &    77.4     \\
 7003.57 & Si~I  &   5.96 &  $-$0.830 &    83.0 &    87.3     \\
 7005.89 & Si~I  &   5.98 &  $-$0.730 &   130.0 &   106.7     \\
 7034.90 & Si~I  &   5.87 &  $-$0.880 &    97.0 &    96.7     \\
 7405.77 & Si~I  &   5.61 &  $-$0.820 &   115.8 &   119.7     \\
 7415.95 & Si~I  &   5.61 &  $-$0.730 &  \nodata &   119.6     \\
 7423.50 & Si~I  &   5.62 &  $-$0.580 &  \nodata &   135.1     \\
 7698.97 & K~I   &   0.00 &  $-$0.168 &   180.0 &   172.1     \\
 5512.99 & Ca~I  &   2.93 &  $-$0.300 &   108.0 &   109.6     \\
 5581.96 & Ca~I  &   2.52 &  $-$0.71  &   117.3 &   113.5     \\
 5590.11 & Ca~I  &   2.52 &  $-$0.710 &   112.2 &   110.8     \\
 5857.45 & Ca~I  &   2.93 &   0.230 &  \nodata &   148.9     \\
 6161.30 & Ca~I  &   2.52 &  $-$1.030 &    80.0 &    75.2     \\
 6166.44 & Ca~I  &   2.52 &  $-$0.90  &    98.6 &    83.8      \\
 6169.04 & Ca~I  &   2.52 &  $-$0.540 &   115.7 &   114.1     \\
 6169.56 & Ca~I  &   2.52 &  $-$0.270 &  \nodata &   134.3     \\
 6471.66 & Ca~I  &   2.52 &  $-$0.590 &   117.9 &   108.8     \\
 6493.78 & Ca~I  &   2.52 &   0.140 &  \nodata &   154.4     \\
 6499.65 & Ca~I  &   2.54 &  $-$0.590 &   108.9 &    98.7     \\
 6508.85 & Ca~I  &   2.52 &  $-$2.120 &  \nodata &    27.2     \\
 6717.68 & Ca~I  &   2.71 &  $-$0.610 &  \nodata &   152.3     \\
 7148.15 & Ca~I  &   2.71 &   0.218 &  \nodata &   169.0     \\
 5526.79 & Sc~II &   1.77 &   0.130 &    86.0 &    92.3     \\
 5657.90 & Sc~II &   1.51 &  $-$0.500 &    89.2 &    79.0     \\
 5667.15 & Sc~II &   1.50 &  $-$1.240 &    66.7 &    58.2     \\
 5669.04 & Sc~II &   1.50 &  $-$1.120 &    60.5 &    56.2     \\
 5684.20 & Sc~II &   1.51 &  $-$1.080 &    57.1 &  \nodata     \\
 6245.64 & Sc~II &   1.51 &  $-$1.130 &    57.7 &    50.8     \\
 6604.60 & Sc~II &   1.36 &  $-$1.31  &    60.3 &    55.1     \\
 5022.87 & Ti~I  &   0.83 &  $-$0.430 &    98.4 &   100.4     \\
 5039.96 & Ti~I  &   0.02 &  $-$1.130 &   107.0 &   106.0     \\
 5210.39 & Ti~I  &   0.05 &  $-$0.880 &   102.8 &  \nodata     \\
 5426.26 & Ti~I  &   0.02 &  $-$3.010 &    13.5 &  \nodata     \\
 5471.20 & Ti~I  &   1.44 &  $-$1.390 &    17.0 &  \nodata     \\
 5490.15 & Ti~I  &   1.46 &  $-$0.933 &    45.1 &  \nodata     \\
 5648.57 & Ti~I  &   2.49 &  $-$0.252 &    33.3 &  \nodata     \\
 5662.16 & Ti~I  &   2.32 &  $-$0.109 &    46.8 &  \nodata     \\
 5689.49 & Ti~I  &   2.30 &  $-$0.469 &    31.4 &    25.6     \\
 5702.69 & Ti~I  &   2.29 &  $-$0.572 &    17.2 &  \nodata     \\
 5739.46 & Ti~I  &   2.25 &  $-$0.602 &    16.9 &  \nodata     \\
 5739.98 & Ti~I  &   2.24 &  $-$0.671 &    13.5 &  \nodata     \\
 5766.33 & Ti~I  &   3.29 &   0.360 &    27.6 &  \nodata     \\
 5866.45 & Ti~I  &   1.07 &  $-$0.840 &    77.6 &    63.6     \\
 5880.27 & Ti~I  &   1.05 &  $-$2.050 &    18.4 &  \nodata     \\
 5903.32 & Ti~I  &   1.07 &  $-$2.140 &    14.3 &  \nodata     \\
 5922.11 & Ti~I  &   1.05 &  $-$1.470 &    45.5 &    46.9     \\
 5937.81 & Ti~I  &   1.07 &  $-$1.890 &    22.0 &  \nodata     \\
 5941.75 & Ti~I  &   1.05 &  $-$1.520 &    45.9 &  \nodata     \\
 5953.16 & Ti~I  &   1.89 &  $-$0.329 &  \nodata &    52.7     \\
 5965.83 & Ti~I  &   1.88 &  $-$0.409 &    62.5 &    52.5     \\
 5978.54 & Ti~I  &   1.87 &  $-$0.496 &    54.0 &    27.4     \\
 6064.63 & Ti~I  &   1.05 &  $-$1.940 &    25.5 &    26.6     \\
 6091.17 & Ti~I  &   2.27 &  $-$0.423 &    32.8 &    23.4     \\
 6092.80 & Ti~I  &   1.89 &  $-$1.380 &     9.7 &  \nodata     \\
 6126.22 & Ti~I  &   1.07 &  $-$1.420 &    46.6 &    33.6     \\
 6258.10 & Ti~I  &   1.44 &  $-$0.355 &    76.4 &    72.1     \\
 6258.71 & Ti~I  &   1.46 &  $-$0.240 &  \nodata &   103.9     \\
 6261.10 & Ti~I  &   1.43 &  $-$0.479 &    82.9 &    68.0     \\
 6303.76 & Ti~I  &   1.44 &  $-$1.570 &    23.0 &  \nodata     \\
 6312.22 & Ti~I  &   1.46 &  $-$1.550 &    22.9 &  \nodata     \\
 6743.12 & Ti~I  &   0.90 &  $-$1.630 &    50.6 &    27.5     \\
 7138.90 & Ti~I  &   1.44 &  $-$1.590 &    18.3 &  \nodata     \\
 7344.69 & Ti~I  &   1.46 &  $-$0.992 &  \nodata &    43.1     \\
 5185.91 & Ti~II &   1.89 &  $-$1.460 &    75.0 &    95.5     \\
 5336.79 & Ti~II &   1.58 &  $-$1.630 &    83.8 &    86.3     \\
 5670.85 & V~I   &   1.08 &  $-$0.425 &    52.2 &    35.0     \\
 5703.57 & V~I   &   1.05 &  $-$0.212 &    65.8 &    65.0     \\
 6081.44 & V~I   &   1.05 &  $-$0.579 &    41.4 &    29.5     \\
 6090.22 & V~I   &   1.08 &  $-$0.062 &    67.4 &    47.2     \\
 6199.20 & V~I   &   0.29 &  $-$1.280 &    38.2 &    22.7     \\
 6243.10 & V~I   &   0.30 &  $-$0.978 &    79.6 &    53.4     \\
 6251.82 & V~I   &   0.29 &  $-$1.340 &    43.0 &    25.5     \\
 6274.64 & V~I   &   0.27 &  $-$1.670 &    23.6 &  \nodata     \\
 6285.14 & V~I   &   0.28 &  $-$1.510 &    37.0 &    23.1     \\
 5345.81 & Cr~I  &   1.00 &  $-$0.970 &  \nodata &   144.3     \\
 5348.33 & Cr~I  &   1.00 &  $-$1.290 &  \nodata &   124.0     \\
 5702.32 & Cr~I  &   3.45 &  $-$0.667 &    43.8 &    36.2     \\
 5783.09 & Cr~I  &   3.32 &  $-$0.500 &    62.5 &    51.5     \\
 5783.89 & Cr~I  &   3.32 &  $-$0.295 &    79.0 &    52.5     \\
 5787.96 & Cr~I  &   3.32 &  $-$0.083 &    69.0 &    64.0     \\
 6979.80 & Cr~I  &   3.46 &  $-$0.411 &    62.0 &    56.4     \\
 5537.74 & Mn~I  &   2.19 &  $-$2.020 &    82.6 &    51.3     \\
 6021.80 & Mn~I  &   3.08 &   0.034 &   125.9 &   119.7     \\
 5198.72 & Fe~I  &   2.22 &  $-$2.140 &   118.6 &   109.4    \\
 5406.78 & Fe~I  &   4.37 &  $-$1.620 &    49.9 &    49.2    \\
 5409.14 & Fe~I  &   4.37 &  $-$1.200 &    85.8 &    65.6    \\
 5417.04 & Fe~I  &   4.41 &  $-$1.580 &    52.9 &    49.8    \\
 5441.33 & Fe~I  &   4.10 &  $-$1.630 &    50.2 &    43.4    \\
 5466.39 & Fe~I  &   4.37 &  $-$0.620 &  \nodata &   122.7    \\
 5473.90 & Fe~I  &   4.15 &  $-$0.690 &  \nodata &    86.7    \\
 5487.14 & Fe~I  &   4.41 &  $-$1.430 &    64.2 &  \nodata    \\
 5494.46 & Fe~I  &   4.07 &  $-$1.990 &    49.3 &  \nodata    \\
 5522.45 & Fe~I  &   4.21 &  $-$1.450 &    62.8 &    56.0    \\
 5525.55 & Fe~I  &   4.23 &  $-$1.080 &    82.6 &    73.1    \\
 5539.29 & Fe~I  &   3.64 &  $-$2.590 &    41.3 &  \nodata    \\
 5554.88 & Fe~I  &   4.55 &  $-$0.350 &   116.4 &   127.3    \\
 5560.21 & Fe~I  &   4.43 &  $-$1.100 &    71.8 &    69.3    \\
 5567.39 & Fe~I  &   2.61 &  $-$2.670 &  \nodata &    85.6    \\
 5568.87 & Fe~I  &   3.63 &  $-$2.850 &    25.0 &  \nodata    \\
 5579.34 & Fe~I  &   4.23 &  $-$2.320 &    25.4 &  \nodata    \\
 5618.63 & Fe~I  &   4.21 &  $-$1.630 &    65.6 &    62.9    \\
 5619.59 & Fe~I  &   4.39 &  $-$1.530 &    63.0 &    45.5    \\
 5620.49 & Fe~I  &   4.15 &  $-$1.810 &  \nodata &    59.9    \\
 5624.04 & Fe~I  &   4.39 &  $-$1.220 &    76.9 &    73.9    \\
 5641.44 & Fe~I  &   4.26 &  $-$1.080 &  \nodata &    71.9    \\
 5650.02 & Fe~I  &   5.10 &  $-$0.820 &    69.5 &  \nodata    \\
 5652.32 & Fe~I  &   4.26 &  $-$1.850 &    41.2 &    32.0    \\
 5653.89 & Fe~I  &   4.39 &  $-$1.540 &    52.3 &    46.0    \\
 5661.35 & Fe~I  &   4.28 &  $-$1.760 &    46.4 &  \nodata    \\
 5662.52 & Fe~I  &   4.18 &  $-$0.570 &   115.6 &   123.0    \\
 5667.52 & Fe~I  &   4.48 &  $-$1.500 &    72.1 &  \nodata    \\
 5679.02 & Fe~I  &   4.65 &  $-$0.820 &    77.6 &    63.0    \\
 5680.24 & Fe~I  &   4.19 &  $-$2.480 &    25.3 &    12.4    \\
 5701.54 & Fe~I  &   2.56 &  $-$2.140 &   107.9 &    95.8    \\
 5705.47 & Fe~I  &   4.30 &  $-$1.360 &    61.0 &    47.7    \\
 5731.76 & Fe~I  &   4.26 &  $-$1.200 &    80.2 &    72.7    \\
 5741.85 & Fe~I  &   4.26 &  $-$1.850 &    57.0 &  \nodata    \\
 5752.04 & Fe~I  &   4.55 &  $-$0.940 &    78.1 &    81.3    \\
 5753.12 & Fe~I  &   4.26 &  $-$0.690 &   101.1 &   106.6    \\
 5760.35 & Fe~I  &   3.64 &  $-$2.390 &    40.8 &    30.6    \\
 5762.42 & Fe~I  &   3.64 &  $-$2.180 &  \nodata &    44.1    \\
 5775.06 & Fe~I  &   4.22 &  $-$1.300 &    80.8 &    77.0    \\
 5778.46 & Fe~I  &   2.59 &  $-$3.430 &    43.0 &  \nodata    \\
 5793.91 & Fe~I  &   4.22 &  $-$1.600 &    55.7 &    52.6    \\
 5805.76 & Fe~I  &   5.03 &  $-$1.490 &    26.2 &    23.4    \\
 5806.72 & Fe~I  &   4.61 &  $-$0.950 &    82.4 &    70.8    \\
 5807.78 & Fe~I  &   3.29 &  $-$3.350 &    23.5 &  \nodata    \\
 5827.88 & Fe~I  &   3.28 &  $-$3.310 &    26.2 &    18.5    \\
 5852.22 & Fe~I  &   4.55 &  $-$1.230 &    58.6 &    57.0    \\
 5855.09 & Fe~I  &   4.61 &  $-$1.480 &    39.9 &    35.5    \\
 5856.08 & Fe~I  &   4.29 &  $-$1.330 &    55.4 &    48.3    \\
 5859.60 & Fe~I  &   4.55 &  $-$0.550 &    89.8 &    85.6    \\
 5862.35 & Fe~I  &   4.55 &  $-$0.330 &   113.2 &   102.7    \\
 5873.21 & Fe~I  &   4.26 &  $-$2.040 &    41.1 &  \nodata    \\
 5881.28 & Fe~I  &   4.61 &  $-$1.740 &    30.2 &    30.1    \\
 5883.81 & Fe~I  &   3.96 &  $-$1.260 &    88.7 &    79.5    \\
 5927.79 & Fe~I  &   4.65 &  $-$0.990 &    58.4 &    49.4    \\
 5929.67 & Fe~I  &   4.55 &  $-$1.310 &    60.8 &    33.5    \\
 5930.17 & Fe~I  &   4.65 &  $-$0.140 &   112.7 &   111.1    \\
 5934.65 & Fe~I  &   3.93 &  $-$1.070 &    97.0 &    86.9    \\
 5940.99 & Fe~I  &   4.18 &  $-$2.050 &    34.3 &    40.1    \\
 5952.72 & Fe~I  &   3.98 &  $-$1.340 &    92.2 &    90.6    \\
 5956.69 & Fe~I  &   0.86 &  $-$4.500 &    73.8 &    77.6    \\
 5976.79 & Fe~I  &   3.94 &  $-$1.330 &    89.8 &    81.4    \\
 5983.69 & Fe~I  &   4.55 &  $-$0.660 &    89.3 &    89.4    \\
 5984.83 & Fe~I  &   4.73 &  $-$0.260 &   111.1 &   105.5    \\
 6024.05 & Fe~I  &   4.55 &   0.030 &  \nodata &   125.9    \\
 6027.05 & Fe~I  &   4.07 &  $-$1.090 &    82.6 &    76.9    \\
 6055.99 & Fe~I  &   4.73 &  $-$0.370 &    92.4 &    86.4    \\
 6078.50 & Fe~I  &   4.79 &  $-$0.330 &   105.0 &   106.1    \\
 6079.00 & Fe~I  &   4.65 &  $-$1.020 &    66.4 &    56.0    \\
 6089.57 & Fe~I  &   5.02 &  $-$0.900 &    57.5 &    54.3    \\
 6093.67 & Fe~I  &   4.61 &  $-$1.400 &    50.6 &    45.0    \\
 6094.37 & Fe~I  &   4.65 &  $-$1.840 &    38.1 &    24.9    \\
 6096.66 & Fe~I  &   3.98 &  $-$1.830 &    59.5 &    54.5    \\
 6151.62 & Fe~I  &   2.18 &  $-$3.370 &    72.2 &    81.2    \\
 6157.73 & Fe~I  &   4.07 &  $-$1.160 &    89.3 &    70.2    \\
 6159.37 & Fe~I  &   4.61 &  $-$1.920 &    24.9 &  \nodata    \\
 6165.36 & Fe~I  &   4.14 &  $-$1.470 &    65.0 &    57.7    \\
 6173.34 & Fe~I  &   2.22 &  $-$2.880 &    93.6 &    77.7    \\
 6180.20 & Fe~I  &   2.73 &  $-$2.650 &    91.6 &    84.0    \\
 6187.99 & Fe~I  &   3.94 &  $-$1.620 &    66.2 &    72.5    \\
 6200.31 & Fe~I  &   2.61 &  $-$2.370 &    97.2 &    98.0    \\
 6240.65 & Fe~I  &   2.22 &  $-$3.170 &    70.8 &    65.2    \\
 6265.13 & Fe~I  &   2.18 &  $-$2.540 &   114.9 &   108.1    \\
 6271.28 & Fe~I  &   3.33 &  $-$2.700 &    50.3 &  \nodata    \\
 6297.79 & Fe~I  &   2.22 &  $-$2.640 &    99.7 &  \nodata    \\
 6302.50 & Fe~I  &   3.69 &  $-$1.110 &   116.2 &   104.3    \\
 6315.81 & Fe~I  &   4.07 &  $-$1.610 &    65.0 &    48.5    \\
 6355.03 & Fe~I  &   2.84 &  $-$2.290 &  \nodata &    93.8    \\
 6380.75 & Fe~I  &   4.19 &  $-$1.380 &    74.3 &    70.1    \\
 6392.54 & Fe~I  &   2.28 &  $-$3.990 &    32.2 &    42.2    \\
 6408.03 & Fe~I  &   3.69 &  $-$1.020 &  \nodata &   113.9    \\
 6469.21 & Fe~I  &   4.83 &  $-$0.730 &    84.3 &    80.8    \\
 6475.63 & Fe~I  &   2.56 &  $-$2.940 &    90.2 &    75.5    \\
 6481.87 & Fe~I  &   2.28 &  $-$3.010 &    87.5 &    93.3    \\
 6495.74 & Fe~I  &   4.83 &  $-$0.840 &    67.7 &    67.5    \\
 6498.94 & Fe~I  &   0.96 &  $-$4.690 &    74.1 &    58.0    \\
 6533.93 & Fe~I  &   4.56 &  $-$1.360 &    59.2 &    54.6    \\
 6546.24 & Fe~I  &   2.76 &  $-$1.540 &   129.6 &   126.1    \\
 6581.21 & Fe~I  &   1.48 &  $-$4.680 &    51.9 &    31.8    \\
 6593.87 & Fe~I  &   2.43 &  $-$2.370 &   114.5 &   107.1    \\
 6597.56 & Fe~I  &   4.79 &  $-$0.970 &    65.2 &    61.3    \\
 6608.02 & Fe~I  &   2.28 &  $-$3.930 &    40.2 &    24.2    \\
 6609.11 & Fe~I  &   2.56 &  $-$2.660 &    94.6 &    92.3    \\
 6625.02 & Fe~I  &   1.01 &  $-$5.370 &    44.1 &    29.9    \\
 6627.54 & Fe~I  &   4.55 &  $-$1.580 &    48.7 &    40.0    \\
 6646.93 & Fe~I  &   2.61 &  $-$3.960 &    28.1 &  \nodata    \\
 6648.12 & Fe~I  &   1.01 &  $-$5.920 &    19.9 &    17.6    \\
 6703.57 & Fe~I  &   2.76 &  $-$3.060 &    56.2 &    49.7    \\
 6713.77 & Fe~I  &   4.79 &  $-$1.500 &    35.6 &    27.7    \\
 6715.38 & Fe~I  &   4.61 &  $-$1.540 &    56.3 &    45.6    \\
 6716.22 & Fe~I  &   4.58 &  $-$1.850 &    35.9 &    31.9    \\
 6725.35 & Fe~I  &   4.19 &  $-$2.250 &    32.0 &    34.3    \\
 6726.67 & Fe~I  &   4.61 &  $-$1.070 &    66.3 &    60.8    \\
 6733.15 & Fe~I  &   4.64 &  $-$1.480 &    47.4 &    48.5    \\
 6739.52 & Fe~I  &   1.56 &  $-$4.790 &    26.4 &  \nodata    \\
 6746.95 & Fe~I  &   2.61 &  $-$4.300 &    12.5 &  \nodata    \\
 6750.15 & Fe~I  &   2.42 &  $-$2.580 &    99.3 &    93.0    \\
 6752.71 & Fe~I  &   4.64 &  $-$1.200 &    59.9 &    47.8    \\
 6786.86 & Fe~I  &   4.19 &  $-$1.970 &    57.1 &  \nodata    \\
 6837.02 & Fe~I  &   4.59 &  $-$1.690 &    32.5 &    27.3    \\
 6839.83 & Fe~I  &   2.56 &  $-$3.350 &    56.7 &    60.7    \\
 6842.68 & Fe~I  &   4.64 &  $-$1.220 &    65.8 &    57.3    \\
 6843.65 & Fe~I  &   4.55 &  $-$0.830 &    84.8 &    74.8    \\
 6855.18 & Fe~I  &   4.56 &  $-$0.740 &    96.7 &    89.9    \\
 6855.71 & Fe~I  &   4.61 &  $-$1.780 &    39.9 &    28.4    \\
 6858.15 & Fe~I  &   4.61 &  $-$0.930 &    70.8 &    72.7    \\
 6861.95 & Fe~I  &   2.42 &  $-$3.850 &    46.0 &  \nodata    \\
 6862.49 & Fe~I  &   4.56 &  $-$1.470 &    49.8 &    39.1    \\
 6971.93 & Fe~I  &   3.02 &  $-$3.340 &  \nodata &    20.0    \\
 6978.85 & Fe~I  &   2.48 &  $-$2.450 &   103.6 &    97.2    \\
 6988.52 & Fe~I  &   2.40 &  $-$3.560 &    59.4 &    52.6    \\
 6999.88 & Fe~I  &   4.10 &  $-$1.460 &    76.1 &    73.4    \\
 7000.62 & Fe~I  &   4.14 &  $-$2.390 &    36.7 &    34.5    \\
 7007.96 & Fe~I  &   4.18 &  $-$1.960 &    46.6 &    41.5    \\
 7014.98 & Fe~I  &   2.45 &  $-$4.200 &    17.6 &  \nodata    \\
 7022.95 & Fe~I  &   4.19 &  $-$1.150 &    88.2 &    85.5    \\
 7038.22 & Fe~I  &   4.22 &  $-$1.200 &   103.5 &    86.7    \\
 7107.46 & Fe~I  &   4.19 &  $-$2.040 &    49.7 &    41.3    \\
 7112.17 & Fe~I  &   2.99 &  $-$3.000 &    63.8 &    46.1    \\
 7114.55 & Fe~I  &   2.69 &  $-$4.000 &    22.7 &  \nodata    \\
 7130.92 & Fe~I  &   4.22 &  $-$0.750 &   125.8 &   113.2    \\
 7132.98 & Fe~I  &   4.07 &  $-$1.630 &    63.4 &    53.4    \\
 7142.52 & Fe~I  &   4.95 &  $-$1.030 &    60.9 &    49.9    \\
 7151.47 & Fe~I  &   2.48 &  $-$3.660 &    59.0 &    39.4    \\
 7181.20 & Fe~I  &   4.22 &  $-$1.250 &  \nodata &    70.4    \\
 7284.84 & Fe~I  &   4.14 &  $-$1.700 &    61.0 &  \nodata    \\
 7285.27 & Fe~I  &   4.61 &  $-$1.660 &    42.2 &  \nodata    \\
 7306.56 & Fe~I  &   4.18 &  $-$1.690 &    66.8 &  \nodata    \\
 7401.69 & Fe~I  &   4.19 &  $-$1.350 &    64.0 &    56.6    \\
 7411.16 & Fe~I  &   4.28 &  $-$0.280 &  \nodata &   121.6    \\
 7418.67 & Fe~I  &   4.14 &  $-$1.380 &    71.3 &    64.2    \\
 7440.92 & Fe~I  &   4.91 &  $-$0.720 &    84.6 &    75.7    \\
 7443.02 & Fe~I  &   4.19 &  $-$1.780 &    64.5 &  \nodata    \\
 7447.40 & Fe~I  &   4.95 &  $-$1.090 &    56.3 &    43.0    \\
 7454.00 & Fe~I  &   4.19 &  $-$2.370 &    32.4 &    23.0    \\
 7461.52 & Fe~I  &   2.56 &  $-$3.530 &    54.5 &    43.3    \\
 7491.65 & Fe~I  &   4.30 &  $-$1.070 &    90.7 &    81.2    \\
 7498.53 & Fe~I  &   4.14 &  $-$2.220 &    36.7 &    21.3    \\
 7568.91 & Fe~I  &   4.28 &  $-$0.940 &   102.2 &    93.8    \\
 7583.79 & Fe~I  &   3.02 &  $-$1.890 &   112.1 &   101.3    \\
 7588.31 & Fe~I  &   5.03 &  $-$1.210 &    55.0 &    46.1    \\
 7751.12 & Fe~I  &   4.99 &  $-$0.850 &    73.3 &    66.8    \\
 7807.92 & Fe~I  &   4.99 &  $-$0.620 &    86.9 &    78.5    \\
 5197.58 & Fe~II &   3.23 &  $-$2.230 &    90.3 &  \nodata    \\
 5234.63 & Fe~II &   3.22 &  $-$2.220 &    97.5 &   101.0    \\
 5414.08 & Fe~II &   3.22 &  $-$3.620 &    39.9 &  \nodata    \\
 5425.26 & Fe~II &   3.00 &  $-$3.240 &    58.9 &    61.5    \\
 5534.85 & Fe~II &   3.25 &  $-$2.640 &    71.0 &    88.4    \\
 5991.38 & Fe~II &   3.15 &  $-$3.570 &    46.5 &    60.1    \\
 6084.11 & Fe~II &   3.20 &  $-$3.800 &    28.8 &    36.0    \\
 6149.26 & Fe~II &   3.89 &  $-$2.690 &    48.2 &    48.7    \\
 6247.56 & Fe~II &   3.89 &  $-$2.360 &    65.2 &    77.2    \\
 6369.46 & Fe~II &   2.89 &  $-$4.200 &    28.3 &    34.2    \\
 6416.92 & Fe~II &   3.89 &  $-$2.690 &    46.3 &    48.1    \\
 6516.08 & Fe~II &   2.89 &  $-$3.450 &    64.6 &    73.7    \\
 7449.34 & Fe~II &   3.89 &  $-$3.310 &    37.1 &    33.0    \\
 5530.79 & Co~I  &   1.71 &  $-$2.060 &    49.8 &    35.1     \\
 5647.23 & Co~I  &   2.28 &  $-$1.560 &    34.4 &  \nodata     \\
 6189.00 & Co~I  &   1.71 &  $-$2.450 &    25.3 &    20.4     \\
 6632.45 & Co~I  &   2.28 &  $-$2.000 &    27.4 &    12.9     \\
 7417.41 & Co~I  &   2.04 &  $-$2.070 &    29.7 &    16.9     \\
 5578.72 & Ni~I  &   1.68 &  $-$2.640 &    83.9 &    66.5     \\
 5587.86 & Ni~I  &   1.93 &  $-$2.140 &  \nodata &    75.3     \\
 5589.36 & Ni~I  &   3.90 &  $-$1.140 &    43.0 &    36.6     \\
 5593.74 & Ni~I  &   3.90 &  $-$0.840 &    66.4 &    61.2     \\
 5625.32 & Ni~I  &   4.09 &  $-$0.701 &    61.4 &    52.8     \\
 5682.20 & Ni~I  &   4.10 &  $-$0.469 &    78.6 &    70.5     \\
 5748.35 & Ni~I  &   1.68 &  $-$3.260 &  \nodata &    38.3     \\
 5760.83 & Ni~I  &   4.10 &  $-$0.805 &    60.7 &    64.9     \\
 5796.09 & Ni~I  &   1.95 &  $-$3.690 &    27.3 &  \nodata     \\
 5805.22 & Ni~I  &   4.17 &  $-$0.638 &    60.1 &    53.7     \\
 5846.99 & Ni~I  &   1.68 &  $-$3.210 &    46.4 &    48.0     \\
 6053.69 & Ni~I  &   4.23 &  $-$1.070 &    41.4 &    42.9     \\
 6086.28 & Ni~I  &   4.26 &  $-$0.515 &    65.7 &    55.5     \\
 6128.97 & Ni~I  &   1.68 &  $-$3.330 &    47.3 &    34.3     \\
 6130.13 & Ni~I  &   4.26 &  $-$0.959 &    41.6 &    36.0     \\
 6175.37 & Ni~I  &   4.09 &  $-$0.535 &    76.5 &    70.4     \\
 6176.81 & Ni~I  &   4.09 &  $-$0.529 &    86.0 &    79.4     \\
 6177.24 & Ni~I  &   1.83 &  $-$3.510 &    38.7 &    30.8     \\
 6186.71 & Ni~I  &   4.10 &  $-$0.965 &    56.4 &    57.4     \\
 6204.60 & Ni~I  &   4.09 &  $-$1.140 &    47.0 &    35.8     \\
 6314.66 & Ni~I  &   1.93 &  $-$1.770 &   114.2 &    99.1     \\
 6360.82 & Ni~I  &   4.17 &  $-$1.150 &    34.1 &    35.8     \\
 6370.35 & Ni~I  &   3.54 &  $-$1.940 &  \nodata &    25.7     \\
 6378.25 & Ni~I  &   4.15 &  $-$0.899 &    55.1 &    52.7     \\
 6482.80 & Ni~I  &   1.93 &  $-$2.630 &    70.2 &    72.8     \\
 6586.31 & Ni~I  &   1.95 &  $-$2.810 &    69.4 &    63.8     \\
 6598.60 & Ni~I  &   4.23 &  $-$0.978 &    49.3 &    47.0     \\
 6635.12 & Ni~I  &   4.42 &  $-$0.828 &  \nodata &    34.7     \\
 6643.63 & Ni~I  &   1.68 &  $-$2.300 &   127.8 &   109.2     \\
 6767.77 & Ni~I  &   1.83 &  $-$2.170 &   103.7 &   100.8     \\
 6772.31 & Ni~I  &   3.66 &  $-$0.987 &    68.2 &    74.8     \\
 6842.04 & Ni~I  &   3.66 &  $-$1.470 &    52.7 &    54.1     \\
 7030.01 & Ni~I  &   3.54 &  $-$1.730 &    34.7 &    33.6     \\
 7110.88 & Ni~I  &   1.93 &  $-$2.970 &    80.2 &    59.6     \\
 7122.20 & Ni~I  &   3.54 &   0.048 &  \nodata &   142.5     \\
 7414.50 & Ni~I  &   1.99 &  $-$2.570 &    98.0 &    93.2     \\
 7422.27 & Ni~I  &   3.63 &  $-$0.129 &   128.7 &   124.4     \\
 7574.05 & Ni~I  &   3.83 &  $-$0.580 &    92.3 &    88.6     \\
 7727.62 & Ni~I  &   3.68 &  $-$0.162 &   114.1 &   115.4     \\
 7748.89 & Ni~I  &   3.70 &  $-$0.130 &   116.7 &   113.5     \\
 7788.93 & Ni~I  &   1.95 &  $-$2.420 &  \nodata &   119.5     \\
 7797.59 & Ni~I  &   3.90 &  $-$0.180 &   105.6 &   100.1     \\
 7826.77 & Ni~I  &   3.70 &  $-$1.950 &    27.7 &    27.7     \\
 5105.54 & Cu~I  &   1.39 &  $-$1.505 &   138.0 &   124.8     \\
 5782.12 & Cu~I  &   1.64 &  $-$1.780 &   135.9 &   125.8     \\
 6362.34 & Zn~I  &   5.79 &   0.140 &    33.0 &    33.5     \\
 5853.70 & Ba~II &   0.60 &  $-$1.010 &    76.0 &    62.2     \\
 6141.70 & Ba~II &   0.70 &  $-$0.070 &   127.0 &   125.9     \\
 6496.90 & Ba~II &   0.60 &  $-$0.380 &   110.5 &   111.1     \\
 4883.69 & Y~II  &   1.08 &   0.070 &    75.0 &  \nodata     \\
 5087.43 & Y~II  &   1.08 &  $-$0.170 &    65.3 &    48.0     \\
 5200.42 & Y~II  &   0.99 &  $-$0.570 &    45.0 &  \nodata     \\
 6127.44 & Zr~I  &   0.15 &  $-$1.060 &     8.0 &  \nodata     \\
 6134.55 & Zr~I  &   0.00 &  $-$1.280 &     5.5 &  \nodata     \\
 5319.81 & Nd~II &   0.55 &  $-$0.140 &    18.0 &  \nodata     \\
\enddata
\end{deluxetable}

\clearpage

\begin{deluxetable}{lllr r | rr | c}
\tabletypesize{\footnotesize}
\tablenum{5}
\tablewidth{0pt}
\tablecaption{Abundances in \moaten
\label{table_abunds_moa310}}
\tablehead{
\colhead{Species} & \colhead{log[$\epsilon(X)$]\tablenotemark{a}} & 
\colhead{$\sigma_{obs}$\tablenotemark{b}} & \colhead{Num. of} &
\colhead{log[$\epsilon(X)/\epsilon(X)_{\odot}]$} &
\colhead{[X/Fe]\tablenotemark{k}} &
\colhead{$\sigma_{pred}$~for} &
 \colhead{Notes}  \\
\colhead{} & \colhead{(dex)} &  \colhead{(dex)} & \colhead{Lines} &
   \colhead{(dex)} & colhead{(dex)} & \colhead{[X/Fe] (dex)} 
 }
\startdata 
C(CH) & 8.89 & 0.15 & band  & +0.30 & $-0.10$ & 0.17  & syn  \\
O~I & 9.09   &  0.16 & 4 & +0.20 & $-0.22$ &  0.19 & high $\chi$ \\
Na~I & 6.63  &  0.16 & 4 & +0.54 &     +0.12 &    0.09 \\
Mg~I & 8.06  &  0.17 & 3 & +0.59 &     +0.17 &    0.07 \\
Al~I & 6.72 &  0.15 & 2 & +0.54 &     +0.12 &    0.08 \\
Si~I & 8.05 &  0.17 & 15 & +0.52 &     +0.10 &    0.17 & high $\chi$ \\
K~I & 5.45  & \nodata & 1 & +0.22 &   $-$0.20 &    0.12 \\
Ca~I & 6.45  & 0.12 & 8 & +0.34 &     $-0.08$ &   0.07    \\
Sc~II & 3.73 &  0.15 & 6 & +0.50 &  +0.08 &    0.10 & d \\
Ti~I & 5.31  &  0.13 & 31  & +0.45  &  +0.03 &   0.11 \\
Ti~II & 5.32 &  0.04 & 2 & +0.45 &      +0.03 & 0.10  \\
V~I   & 4.40 &  0.10 & 9 & +0.61 &     +0.19 &    0.14 & d \\
Cr~I & 6.17 &  0.15 & 5 & +0.51 &     +0.09 &    0.07 \\
Mn~I & 5.79 &  0.09 & 2 & +0.42 &     +0.00 &    0.11 & e \\
Fe~I &  7.90 &  0.14 & 100 & +0.42 &  0.00  & 0.09\tablenotemark{i}  \\
Fe~II & 7.87 &  0.14 &  13 & +0.39   & $-$0.03 & 0.17\tablenotemark{j} \\
Co~I  & 5.40 &  0.11 &  5 & +0.63 &     +0.19 &    0.08 & d \\
Ni~I  & 6.74 &  0.15 &  37  & +0.56 &     +0.14 &    0.05 \\
Cu~I & 4.78 & 0.20 & 2 & +0.81 &     +0.39 &    0.15 & f \\
Zn~I & 4.92 &  \nodata & 1 & +0.37 &  $-0.05$ &    0.13 \\
Y~II & 2.59 & 0.15 & 3 & +0.53 & 0.11 & 0.12 \\
Ba~II & 2.60 &  0.04 & 3 & +0.31 &  $-$0.11 &    0.17 & d \\
Nd~II & 1.77 & \nodata & 1 & +0.31 & $-0.11$ & 0.12 \\
\enddata
\tablenotetext{a}{This is log[$(n(X)/n(H)$] + 12.0~dex.}
\tablenotetext{b}{Rms dispersion about the mean abundance, using 
differential line-by-line abundances with respect to the Sun.}
\tablenotetext{d}{The HFS corrections are small and not an issue.}
\tablenotetext{e}{The HFS corrections are large and are a concern.}
\tablenotetext{f}{The HFS corrections are very large and are a major concern.}
\tablenotetext{i}{The uncertainty in [Fe/H] inferred from the 100 Fe~I lines.}
\tablenotetext{j}{The uncertainty in [Fe/H] inferred from the 13 Fe~II lines.}
\tablenotetext{k}{The reference species (Fe~I or Fe~II) is 
based on the level of excitation and ionization.  See Table~4
in \cite{cohen08}.}
\end{deluxetable}

\begin{deluxetable}{lllr r | rr | c}
\tabletypesize{\footnotesize}
\tablenum{6}
\tablewidth{0pt}
\tablecaption{Abundances in \moaeleven
\label{table_abunds_moa311}}
\tablehead{
\colhead{Species} & \colhead{log[$\epsilon(X)$]\tablenotemark{a}} & 
\colhead{$\sigma_{obs}$\tablenotemark{b}} & \colhead{Num. of} &
\colhead{log[$\epsilon(X)/\epsilon(X)_{\odot}]$} &
\colhead{[X/Fe]\tablenotemark{k}} &
\colhead{$\sigma_{pred}$~for} &
 \colhead{Notes}  \\
\colhead{} & \colhead{(dex)} &  \colhead{(dex)} & \colhead{Lines} &
   \colhead{(dex)} & colhead{(dex)} & \colhead{[X/Fe] (dex)} 
 }
\startdata 
C(CH) & 8.89 & 0.15 & band  & +0.30 & +0.05 & 0.17  & syn  \\
O~I & 9.31   &  0.12 & 4 & +0.42 & +0.14 &  0.19 & high $\chi$ \\
Na~I & 6.54  &  0.09 & 4 & +0.45 &     +0.20 &    0.09 \\
Mg~I & 7.91  &  0.14 & 4 & +0.33 &     +0.08 &    0.07 \\ 
Al~I & 6.72 &  0.09 & 2 & +0.54 &     +0.29 &    0.08 \\
Si~I & 7.94 &  0.12 & 15 & +0.41 &     +0.16 &    0.17 & high $\chi$ \\
K~I & 5.45  & \nodata & 1 & +0.22 &   $-$0.03 &    0.12 \\
Ca~I & 6.49  & 0.18 & 9 & +0.36 &     +0.11 &   0.07    \\
Sc~II & 3.47 &  0.11 & 6 & +0.24 &    $-0.04$ &    0.10 & d \\
Ti~I & 5.25  &  0.18 & 15  & +0.42  &  +0.17 &   0.11 \\
Ti~II & 5.30 &  0.12 & 2 & +0.43 &      +0.15 & 0.10  \\
V~I   & 4.10 &  0.10 & 8 & +0.31 &     +0.06 &    0.14 & d \\
Cr~I & 5.98 &  0.11 & 6 & +0.32 &     +0.07 &    0.07 \\
Mn~I & 5.66 &  0.12 & 2 & +0.29 &     +0.04 &    0.11 & e \\
Fe~I &  7.73 &  0.16 & 92 & +0.25 &  0.00  & 0.09\tablenotemark{i}  \\
Fe~II & 7.75 &  0.16 &  11 & +0.28   & +0.03 & 0.17\tablenotemark{j} \\
Co~I  & 5.14 &  0.08 &  4 & +0.36 &     +0.11 &    0.08 & d \\
Ni~I  & 6.60 &  0.15 &  41  & +0.42 &     +0.17 &    0.05 \\
Cu~I & 4.35 & 0.09 & 2 & +0.38 &     +0.13 &    0.15 & f \\
Zn~I & 4.84 &  \nodata & 1 & +0.29 &  +0.04 &    0.13 \\
Y~II & 2.34 & \nodata & 1 & +0.45 & 0.17 & 0.12 \\
Ba~II & 2.24 &  0.11 & 3 & +0.12 &  $-$0.13 &    0.17 & d \\
\enddata
\tablenotetext{a}{This is log[$(n(X)/n(H)$] + 12.0~dex.}
\tablenotetext{b}{Rms dispersion about the mean abundance, using 
differential line-by-line abundances with respect to the Sun.}
\tablenotetext{d}{The HFS corrections are small and not an issue.}
\tablenotetext{e}{The HFS corrections are large and are a concern.}
\tablenotetext{f}{The HFS corrections are very large and are a major concern.}
\tablenotetext{i}{The uncertainty in [Fe/H] inferred from the 92 Fe~I lines.}
\tablenotetext{j}{The uncertainty in [Fe/H] inferred from the 11 Fe~II lines.}
\tablenotetext{k}{The reference species (Fe~I or Fe~II) is 
based on the level of excitation and ionization.  See Table~4
in \cite{cohen08}.}
\end{deluxetable}

\begin{deluxetable}{l r}
\tablenum{7}
\tablewidth{0pt}
\tablecaption{Probability for Identical Fe-Metallicity Distributions for 
the 6 Microlensed Dwarfs
and the Bulge Giants
\label{table_prob}}
\tablehead{
\colhead{Systematic Offset\tablenotemark{a}} & 
\colhead{Prob. $<$[Fe/H]$>$(6 Dwarfs)\tablenotemark{b}} \\  
\colhead{(dex)} & \colhead{(\%)} 
}
\startdata
0.0 & 0.39  \\
$-0.05$ & 1.65  \\
$-0.10$ & 4.94 \\
$-0.15$ & 11.53  \\
$-0.20$ & 21.30  \\
\enddata
\tablenotetext{a}{The systematic offset between the Fe-metallicity
scale of \cite{zoccali08} and that for the abundances of the
6 microlensed main sequent turnoff region stars in the Galactic bulge.}
\tablenotetext{b}{The probability of achieving the mean [Fe/H]
for the 6 dwarfs, +0.29~dex, from the \cite{zoccali08} Fe-metallicity 
distribution function for Baade's Window.}
\end{deluxetable}

\clearpage

\clearpage

\begin{thebibliography}{}






\bibitem[Bensby \etal(2005)]{bensby05}
Bensby,~T., Feltzing,~S., Lundstrom,~I. \& Ilyin, I., 2005, \aap, 433, 185

\bibitem[Bensby \etal(2009)]{bensby09}
Bensby, T. \etal, 2009, \aap, in press

\bibitem[Bernstein et al(2003)]{bernstein03}
Bernstein, R., Shectman, S.~A., Gunnels, S.~M., Mochnacki, S.
\& Athey, A.~E., 2003, SPIE, 4841, {\it{Instrument Design and Performance
for Optical/Infrared Ground-Based Telescopes}}, ed. I.~Masanori
\& A.~Moorhead, 1694

\bibitem[Biazzo, Frasca, Catalano \& Marilli(2007)]{biazzo07}
Biazzo,~K., Frasca,~A., Catalano~S. \& Marilli~E., 2007, 
Astr.~Nach., 328, 938

\bibitem[Boesgaard, Jensen \& Deliyannis(2009)]{ngc6791}
Boesgaard, A.~M., Jensen, E.~E.~C. \& Deliyannis, C., 2009, \aj, 137, 4949 






\bibitem[Castelli \& Kurucz(2003)]{no_over}
Castelli, F. \& Kurucz, R.~L., 2003, in Poster Paper A20, 
on CD from IAU Sym. 210,
{\it{Modeling of Stellar Atmospheres}}, ed. N.~E.~Piskunov, 
W.~W.~Weiss \& D.~G.~Gray (San Francisco: ASP) (see Astro-ph/0405087)


\bibitem[Clarkson \etal(2008)]{clarkson08}
Clarkson, W. \etal, 2009, \apj, 684, 1110


\bibitem[Cohen \etal(1999)]{cohen99}
Cohen, J.~G., Gratton, R.~G., Behr, B. \& Carretta, E., 1999, \apj, 523, 739



\bibitem[Cohen \etal(2008)]{cohen08}
Cohen, J.~G.,, Huang, W., Udalski, A.,
Gould, A. \& Johnson, J.~A., 2008, \apj,  682, 1029

\bibitem[Cunha \& Smith(2006)]{cunha06}
Cunha, K. \& Smith, V.~V., 2006, \apj, 651, 491

\bibitem[Dotter \etal(2008)]{dartmouth}
Dotter, A., Chaboyer, B., Jevermovic, D., Kostov, V.,
Baron, E. \& Ferguson, J.~W., 2008, \apjs, 178, 89


\bibitem[Feltzing \& Gilmore(2000)]{feltzing00}
Feltzing, S. \& GIlmore, G., 2000, \aap, 355, 949


\bibitem[Figer et al(2002)]{figer02}
Figer, D.~F. et al, 2002, \apj, 581, 258

\bibitem[Fulbright, McWilliam \& Rich(2006)]{fulbright06}
Fulbright, J.~P., McWilliam, A. \& Rich, R.~M., 2006, \apj, 636, 821


\bibitem[Fulbright, McWilliam \& Rich(2007)]{fulbright07}
Fulbright, J.~P., McWilliam, A. \& Rich, R.~M., 2007, \apj, 661, 1162



\bibitem[Gray \& Johanson(1991)]{gray91}
Gray,~D.F. \& Johanson,~H.~L., 1991, \apj, 103, 439

\bibitem[Grenon(1999)]{grenon99}
Grenon, M., 1999, Astrophysics and Space Science, 265, 331

\bibitem[Holmberg, Flynn \& Portinari(2006)]{holmberg06}
Holmberg,~J., Flynn,~C. \& Portinari, L., 2006, \mnras, 367, 449




\bibitem[Johnson \etal(2007)]{johnson07}
Johnson,~J.~A., Gal-Yam,~A., Leonard,~D.~C., Simon,~J.~D.,
Udalski, ~A. \& Gould,~A., 2007, \apjl, 655, L3

\bibitem[Johnson \etal(2008)]{johnson08}
Johnson,~J.~A., Gaido. B.S., Sumi, T., Bond, I.~A.
\& Gould,~A., 2008, \apj, 685, 508





\bibitem[Kurucz(1993)]{kurucz93} Kurucz, R. L., 1993, ATLAS9 Stellar 
Atmosphere Programs and 2 km/s Grid, (Kurucz CD-ROM No. 13)

\bibitem[Lecureur \etal(2007)]{lecureur07}
Lecureur, A., Hill, V., Zoccali, M., Barbuy, B., Gomez, A.,
Minitti, D., Ortolani, S. \& Renzini, A., 2007, \aap,465, 799

\bibitem[Luck, Kovtyukh \& Andrievsky(2006)]{luck06}
Luck, R.~E., Kovtyukh, V.~V. \& Andrievsky, S.~M., 2006, \aj, 132, 902






\bibitem[Mashonkina et al(2004)]{mashonkina04}
Mashonkina, L., Gehren, T., Travaglio, C., Borkova, T., 2004,
\aap, 433, 185






\bibitem[Pasquini \etal(2004)]{pasquini04}
Pasquini, L., Randich, S., Zoccali, M., Hill, V., Charbonnel, C.
\& Nordstrom, B., 2004, \aap, 424, 951

\bibitem[Pompeia, Barbuy \& Grenon(2002)]{pompeia02}
Pompeia, L., Barbuy, B. \& Grenon, M., 2002, \apj, 566, 845




 


\bibitem[Reddy \etal(2003)]{reddy03}
Reddy, B.~E., Tomkink, J., Lambert D.~L. \& Allende Prieto, C.,
2003, \mnras, 340, 304


\bibitem[Rich \& Origlia(2005)]{rich05}
Rich, R.~M. \& Origlia,~L., 2005, \apj, 634, 1293

\bibitem[Rich, Origlia \& Valenti(2007)]{rich07}
Rich, R.~M., Origlia,~L. \& Valenti, E., 2007, \apjl, 665, L119


\bibitem[Santos \etal(2009)]{santos09}
Santos, N.~C., Lovis, C., Pace, G., Melendez, J. \& Naef, D., 2009, 
\aap, 493, 309

\bibitem[Sneden(1973)]{moog} Sneden, C., 1973, Ph.D. thesis, Univ. 
of Texas
















\bibitem[Zoccali \etal(2003)]{zoccali03}
Zoccali, M. \etal, 2003, \aap, 399, 931 




\bibitem[Zoccali \etal(2008)]{zoccali08}
Zoccali, M., Hill, V., Lecureur, A., Barbuy, B., Renzini, A.,
Minitti, D., Gomez, A. \& Ortolani, S., 2008, \aap, 486, 177

\end{thebibliography}
\end{document}